\documentclass[acmsmall,screen]{acmart}
\AtBeginDocument{%
  }

\usepackage{enumitem}
\usepackage{threeparttable}
\usepackage{array}
\usepackage{multirow}
\usepackage{booktabs}
\usepackage{makecell}
\usepackage[linesnumbered,ruled,vlined,noend]{algorithm2e}
\usepackage{subfigure}
\usepackage[most]{tcolorbox}
\usepackage{listings}
\usepackage{xspace}
\usepackage{pifont}

\lstdefinestyle{prompt}{
    basicstyle=\ttfamily\small,
    backgroundcolor=\color{gray!4},
    frame=single,
    framerule=0.4pt,
    rulecolor=\color{gray!55},
    framesep=5pt,
    breaklines=true,
    breakautoindent=false,
    breakindent=0pt,
    breakatwhitespace=true,
    columns=fullflexible,
    keepspaces=true,
    showstringspaces=false,
    captionpos=b,
    aboveskip=6pt,
    belowskip=6pt,
    keywordstyle=\bfseries\color{blue!45!black},
    morekeywords={SYSTEM,USER}
}

\newtcolorbox{resultbox}{
    enhanced,
    colback=gray!5,
    frame hidden,
    boxrule=0pt,
    borderline west={4pt}{0pt}{gray!80}, 
    sharp corners,
    left=6pt,
    right=2pt,
    top=2pt,
    bottom=2pt,
    fontupper=\relax,
    parbox=false
}

\SetKwInput{Input}{Input}
\SetKwInput{Output}{Output}
\SetKwProg{Fn}{Function}{:}{}

\SetCommentSty{mycommfont}

\newcommand{\Code}[1]{\begin{small}\texttt{#1}\end{small}}

\newcommand{\cmark}{\ding{51}}
\newcommand{\xmark}{\text{\ding{55}}}

\makeatletter
\patchcmd\algocf@Vline{\vrule}{\vrule \kern-0.4pt}{}{}
\patchcmd\algocf@Vsline{\vrule}{\vrule \kern-0.4pt}{}{}
\makeatother

\newcommand{\toolname}{\textsc{Bridge}\xspace}

\setcopyright{acmlicensed}
\copyrightyear{2018}
\acmYear{2018}
\acmDOI{XXXXXXX.XXXXXXX}

\acmJournal{JACM}
\acmVolume{37}
\acmNumber{4}
\acmArticle{111}
\acmMonth{8}

\begin{document}

\title{\toolname: Automatically Mining Ecosystem-Scale API Update Mappings and Client Update Instances}

\author{Kai Gao}
\orcid{0000-0002-0942-7890}
\affiliation{%
 \institution{University of Science and Technology Beijing}
 \city{Beijing}
 \country{China}}
\email{kai.gao@ustb.edu.cn}

\author{Yu Sun}
\orcid{0009-0003-4857-5719}
\affiliation{%
 \institution{University of Science and Technology Beijing}
 \city{Beijing}
 \country{China}}
\email{yusun@xs.ustb.edu.cn}

\author{Chang-ai Sun}
\orcid{0000-0003-3696-6176}
\affiliation{%
 \institution{University of Science and Technology Beijing}
 \city{Beijing}
 \country{China}}
\email{casun@ustb.edu.cn}

\begin{abstract}
Library updates are essential for incorporating bug fixes, security patches, and new features, but they often require adapting client code to API changes. 
\textit{API update mappings} that identify relations between legacy and replacement APIs, \textit{version transitions} that these mappings apply, and \textit{client update instances} that capture concrete API call changes are essential for developing and evaluating automated library update techniques. 
Existing library evolution datasets capture different subsets of this information, but none simultaneously provides all three. 
Moreover, these datasets typically cover a small number of third-party libraries, limiting the diversity of API updates they capture. 
Constructing a dataset that provides all three types of information at ecosystem scale is challenging because it requires reliably connecting API update mappings, version transitions, and client update instances across large numbers of library releases and client projects. 

To address this challenge, we present \toolname, a \textit{client-driven} framework that starts from observed client dependency updates, which naturally connect library version transitions with client code changes. 
Built on the World of Code (WoC) infrastructure, \toolname mines candidate update instances from client dependency update commits at scale, validates them using library-side evidence, and then derives API update mappings from validated instances. 
This design grounds each retained mapping in at least one client update instance. 
On a manually annotated ground truth dataset, \toolname achieves 91.6\% precision and 88.7\% recall for Java and 90.1\% precision and 64.0\% recall for Python. 
Applied to WoC V3, \toolname mines 381,661 Java and 277,259 Python client update instances, representing 18,900 and 4,456 API update mappings across 2,557 and 999 libraries, respectively. 
The mined mappings exhibit a pronounced long-tail distribution, with most appearing in only a few client update instances. 
As one application of the dataset, we evaluate four large language models on replacement API recommendation, a key step in library updates. 
The best recommendation accuracy reaches only 37.1\% for Java and 44.4\% for Python, and all evaluated models perform substantially better on frequently observed mappings than on mappings observed in only a few client update instances, highlighting the difficulty current LLMs face in recommending replacements for mappings in the long tail. 
Overall, \toolname provides an automated approach to constructing ecosystem-scale library update datasets that connect what APIs are replaced, between which versions, and how clients perform the updates. 
\end{abstract}




\received{20 February 2007}
\received[revised]{12 March 2009}
\received[accepted]{5 June 2009}

\maketitle

\section{Introduction}\label{s: intro}
Modern software systems depend extensively on third-party libraries for reusable functionality. 
New library versions provide security patches, bug fixes, and new features, making timely library updates an important dependency management practice for software security and quality~\cite{githubBestPractices, sonatypeOptimizingSoftware, scorecard}. 
However, new releases may deprecate or remove APIs~\cite{Sawant2018-ICSE, Wang2020-FSE}, breaking client code~\cite{10.1145/2950290.2950325, 10.1145/3447245} and imposing substantial maintenance effort on developers~\cite{10.1007/s10664-014-9325-9, 10.1007/s10664-017-9521-5}. 
Consequently, developers often adopt the conservative practice of ``if it ain't broke, don't fix it'' and forgo library updates~\cite{10.1007/s10664-017-9521-5, 10.1145/3133956.3134059, 10.1145/3196321.3196341}. 

Automating library updates requires information about \textit{what} APIs should be replaced and \textit{how} client code should be adapted. 
Specifically, \textit{API update mappings} identify relations between legacy and replacement APIs, their applicable \textit{version transitions} specify when these replacements occur, and \textit{client update instances} capture concrete API call changes performed during library updates. 
Such information underpins rule-based techniques that derive and apply code transformation rules~\cite{CocciEvolve, APIfix, 9609172, LibSync, AppEvolve, A3, openrewrite, Meditor}. 
Recent LLM-based techniques similarly incorporate API mappings, update examples, or natural-language instructions into prompts to adapt client code~\cite{GoogleCodeMigration, SQLAlchemy, GUPPY, AmazonQDeveloper}. 
These three forms of information therefore provide useful data for developing and evaluating automated library update techniques. 

\begin{figure}
    \centering
    \includegraphics[width=0.85\linewidth]{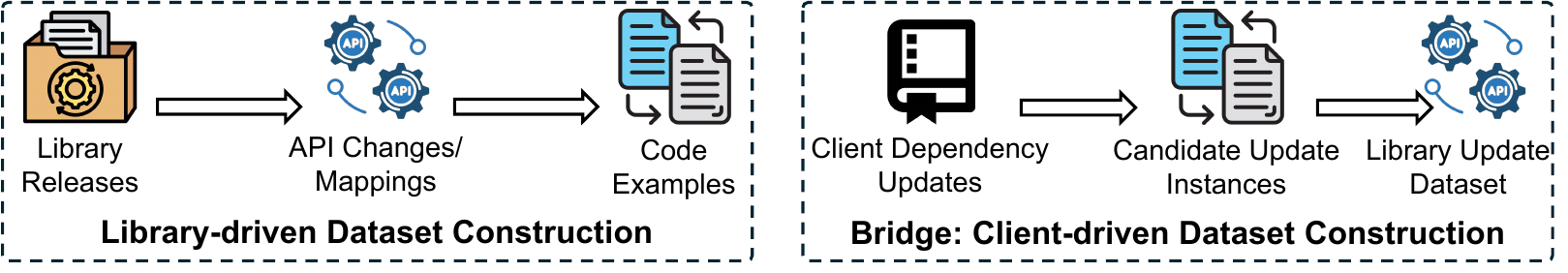}
    \caption{Comparison of library-driven and \toolname's client-driven workflows to constructing library update datasets.}
    \Description{Library-driven dataset construction first identifies API changes or mappings and then prepares corresponding code examples. Bridge instead starts from client dependency updates, mines candidate update instances from associated code changes, and then resolves and validates their underlying API mappings against library artifacts.}
    \label{fig: workflow_comparison}
\end{figure}

Existing library evolution datasets~\cite{Nizar2024-Arxiv, Wu2024-Arxiv, Wang2025-ICSE, Kuhar2025-NAACL, CODEMENV} capture different subsets of this information, but none simultaneously provides all three. 
They also typically cover a small number of third-party libraries, limiting the diversity of API updates they capture. 
As shown in Figure~\ref{fig: workflow_comparison}, their construction generally follows a \textit{library-driven} workflow: they begin with a predefined collection of libraries, identify API changes or mappings within them, and then prepare corresponding code examples from sources such as documentation~\cite{Wu2024-Arxiv, Kuhar2025-NAACL}, client projects~\cite{Wang2025-ICSE}, manual construction~\cite{Nizar2024-Arxiv}, or LLM generation~\cite{CODEMENV}. 
Scaling this workflow is challenging because API update mappings identified from library artifacts must be connected to the version transitions and client code changes in which they are actually observed, which requires analysis across large numbers of library releases and client projects. 
Moreover, client projects typically exercise only a small subset of the APIs exposed by their dependencies~\cite{Huang2022-EMSE, Wang2025-ICSE}, so even a known mapping may have few or no observable client update instances. 
Consequently, \textit{\textbf{ecosystem-scale library update datasets that connect API update mappings, version transitions, and client update instances remain scarce}}. 

To address this challenge, we present \toolname, a \textit{client-driven} framework for automatically constructing library update datasets from observed client dependency updates. 
As illustrated in Figure~\ref{fig: workflow_comparison}, instead of first identifying API update mappings and subsequently obtaining code examples, \toolname starts from dependency update commits in client projects. 
Such commits directly provide the updated library and its old and new versions, while associated source code changes provide candidate adaptations for that version transition. 
\toolname therefore first mines candidate update instances from client-side API call changes, validates the candidates using library-side evidence, and then derives API update mappings from validated instances. 
Its key design principle is that client code provides plausible replacement candidates, while library artifacts provide evidence for validating them. 
Consequently, the resulting library update dataset connects API update mappings, version transitions, and client update instances, with each retained mapping associated with a version transition and grounded in at least one client update instance. 

Built on the World of Code (WoC) infrastructure~\cite{Ma2019-MSR, Ma2021-EMSE}, \toolname realizes this client-driven workflow in four phases. 
It first identifies commits that update dependency versions and also modify source files, which naturally links version transitions and client code changes. 
It then extracts API call metadata from the old and new source blobs. 
Next, \toolname identifies changed calls associated with the updated library and applies similarity-based matching to generate candidate update instances. 
Finally, it validates the candidates against the involved library versions by verifying API existence and the removal or deprecation of the legacy API, and then derives API update mappings from the validated instances. 

We implement \toolname for two popular and substantially different ecosystems, Java and Python, and evaluate it from four perspectives. 
First, we evaluate its accuracy in mining client update instances against a manually annotated ground truth dataset containing 344 and 242 genuine update instances in 638 Java and 609 Python API call change records, respectively. 
\toolname achieves 91.6\% precision, 88.7\% recall, and a 90.1\% F1 score for Java, and 90.1\% precision, 64.0\% recall, and a 74.9\% F1 score for Python. 
Second, we analyze the similarity parameters used for client-side candidate generation and the contribution of library-side validation. 
Similarity weights have limited impact on the F1 score, while the matching threshold controls the precision--recall tradeoff. 
More importantly, library-side validation improves precision by 40.3 percentage points for Java and 43.3 points for Python, empirically supporting the design of combining client-side candidate generation with library-side validation. 
Third, we apply \toolname to WoC V3 and construct a dataset containing 381,661 client update instances and 18,900 API update mappings across 2,557 Java libraries, together with 277,259 instances and 4,456 mappings across 999 Python libraries. 
The mined mappings exhibit a pronounced long-tail distribution: the median mapping appears in only two client update instances, while a small number recur in thousands or tens of thousands of instances. 
Finally, we demonstrate one application of the dataset by evaluating four large language models on replacement API recommendation, a key step in library updates. 
The best recommendation accuracy reaches only 37.1\% for Java and 44.4\% for Python, and all evaluated models perform substantially worse on mappings observed in only a few client update instances. 
By connecting API update mappings with their occurrences in client updates, the dataset enables such frequency-aware analyses and highlights the challenge current LLMs face in recommending replacements for mappings in the long tail. 

In summary, this paper makes the following contributions:
\begin{itemize}[leftmargin=*]
    \item We propose \toolname, a client-driven framework for automatically constructing library update datasets that connect API update mappings, version transitions, and client update instances. 
    \item We implement \toolname for Java and Python and systematically evaluate its effectiveness and key design choices, demonstrating its effectiveness for high-precision library update dataset construction. 
    \item We construct an ecosystem-scale dataset spanning thousands of Java and Python libraries and hundreds of thousands of client update instances, with a pronounced long-tail distribution of API replacements. 
    \item We demonstrate the utility of the dataset through a replacement API recommendation study, revealing that LLMs have substantially greater difficulty in recommending replacements for mappings in the long tail. 
\end{itemize}

The remainder of this paper is organized as follows. 
Section~\ref{s: background} introduces the background and related work, Section~\ref{s: method} presents the \toolname framework, Section~\ref{s: evaluation} reports the evaluation, and Section~\ref{s: discussion} discusses the implications and threats to validity. 
Section~\ref{s: conclusion} concludes the paper. 
The implementation of \toolname and the mined dataset are publicly available at \url{https://github.com/ROCK-SE/Bridge}. 

\section{Background and Related Work}\label{s: background}
This section defines related terminology, reviews existing library evolution datasets and approaches for mining API update mappings and client update instances, and introduces the World of Code infrastructure used by \toolname. 

\subsection{Terminology}\label{ss: terminology}
Many programming language ecosystems provide centralized registries, such as Maven Central in Java and PyPI in Python, for distributing reusable software libraries. 
In this paper, a \textit{library} refers to a third-party project published on such a registry. 
A Java library is identified by its group ID and artifact ID~\cite{pomxml}, such as \Code{junit:junit}, while a Python library is identified by its normalized PyPI name~\cite{pythonNamesNormalization}, such as \Code{django}. 
When a project uses a library, we refer to the library as the project's \textit{dependency} and the project as the library's \textit{client}. 
A library typically releases multiple versions, each exposing a set of public Application Programming Interfaces (APIs). 
We use \textit{API signature} to denote the user-facing identity of an API. 
For Python, we represent an API signature by its fully qualified name (FQN), such as \Code{django.urls.reverse}, following common practice in Python API evolution datasets. 
For Java, an API signature consists of the FQN and formal parameter types~\cite{oracleDefiningMethods}, such as \Code{org.junit.Assert.\allowbreak assertEquals(long, long)}, because Java supports method overloading. 
An \textit{API call} is a concrete invocation of an API in client code, including its actual arguments. 

\begin{figure}
    \centering
    \includegraphics[width=0.85\linewidth]{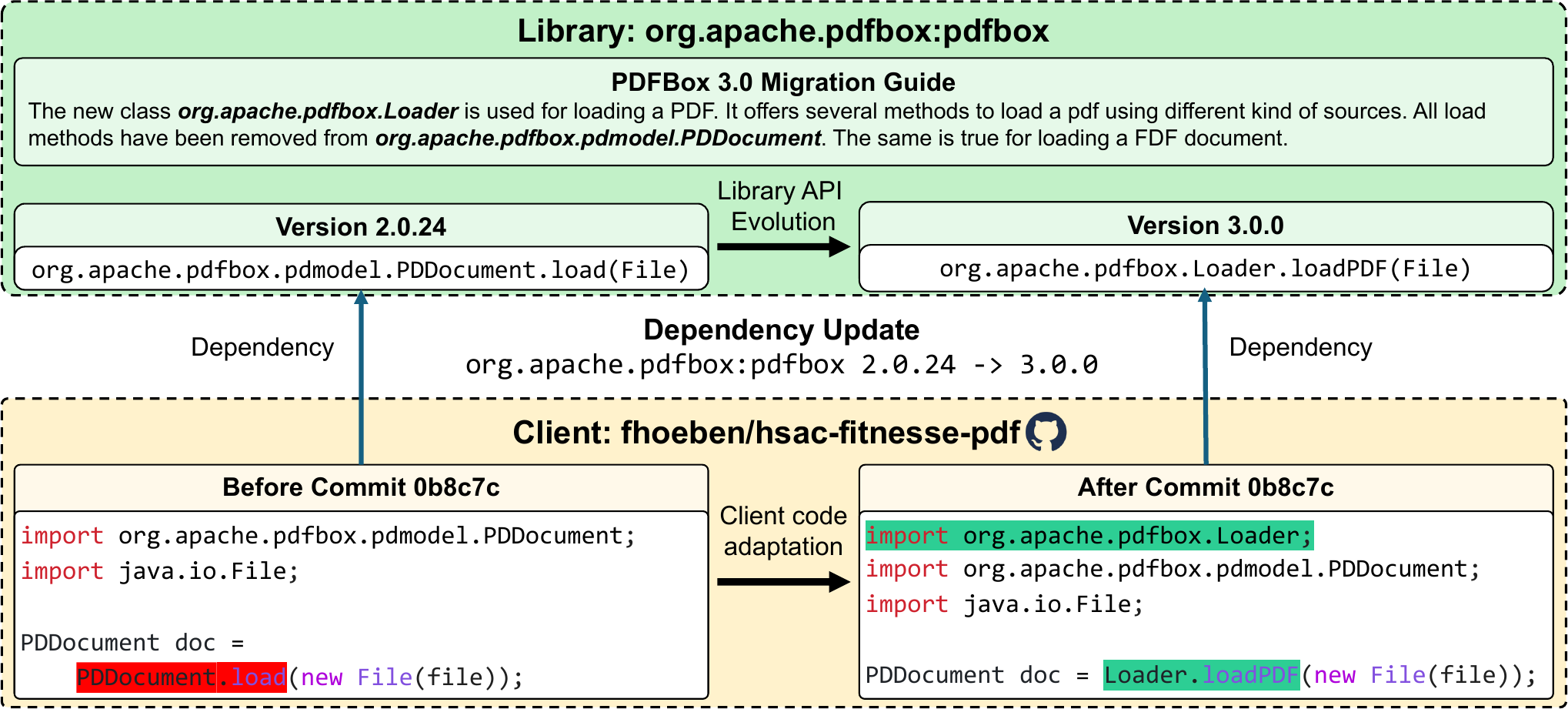}
    \caption{A library update example from commit \href{https://github.com/fhoeben/hsac-fitnesse-pdf/commit/0b8c7c98156809d9bfd5a5dd31d172887c8eefbc}{0b8c7c} in the \href{https://github.com/fhoeben/hsac-fitnesse-pdf}{fhoeben/hsac-fitnesse-pdf} project.}
    \Description{A concrete library update example.}
    \label{fig: update_example}
\end{figure}

When a client updates a dependency from one version to another, its source code may require adaptation to API changes in the updated library. 
Figure~\ref{fig: update_example} shows a concrete example in which a client updates Apache PDFBox from version 2.0.24 to 3.0.0 and replaces the removed \Code{org.apache.pdfbox.pdmodel.PDDocument.load(File)} API with \Code{org.apache.pdfbox.pdmodel.Loader. loadPDF(File)}, consistent with the library's upgrade guide\footnote{\url{https://pdfbox.apache.org/3.0/migration.html\#use-loader-to-get-a-pdf-document}}. 
Prior work has used terms such as API migration~\cite{Meditor, A3, MELT}, code migration~\cite{GoogleCodeMigration, CODEMENV}, and library migration~\cite{LibSync, SQLAlchemy} for such adaptation. 
In this paper, we choose the term \textit{library update}. 
We avoid using ``migration'' for this process because this term is also widely used to denote replacing one library with another that provides similar functionality~\cite{MigrationAdvisor, He2021-FSE, SALM, PyMigBench, PyMigTax, DeepMig, LLM-LM, LibRec, CLM, CMigration}. 

Given a library $l$ and two versions $v_o$ and $v_n$, where $v_o$ denotes the old version and $v_n$ the new version, we define three types of information in our library update dataset, as illustrated by Figure~\ref{fig: update_example}. 
First, an \textbf{API update mapping} is a 3-tuple $(l, a_o, a_n)$, where $a_o$ is a legacy API signature and $a_n$ is its replacement API signature. 
This definition is consistent with prior work~\cite{Wang2025-ICSE}. 
An API update mapping captures a replacement relation independently of a particular version transition. 
For example, Figure~\ref{fig: update_example} shows the mapping from \Code{org.apache.pdfbox.pdmodel.PDDocument.load(File)} to \Code{org.apache.pdfbox.Loader.loadPDF(File)}. 
Second, an \textbf{API update pair} is a 5-tuple $(l, v_o, a_o, v_n, a_n)$ that associates an API update mapping with a particular version transition. 
A single mapping may therefore correspond to multiple update pairs when the same replacement relation applies across different version transitions. 
Third, an \textbf{client update instance} is a 5-tuple $(l, v_o, c_o, v_n, c_n)$, where $c_o$ and $c_n$ are legacy and replacement API calls observed in client code when the client updates $l$ from $v_o$ to $v_n$. 
For example, Figure~\ref{fig: update_example} shows an instance in which the client replaces \Code{PDDocument.load(new File(file))} with \Code{Loader.loadPDF(new File(file))}. 
Thus, mappings capture general replacement relations, update pairs capture their applicability to specific version transitions, and update instances capture how clients perform the replacement in code. 

\subsection{Library Evolution Datasets}\label{ss: library-evolution-datasets}
Automated library update techniques rely on different forms of information about API updates and client code adaptations. 
Rule-based approaches represent code transformation rules using domain-specific languages~\cite{CocciEvolve, APIfix, MELT, 9609172, MLCatchUp}, edit scripts~\cite{LibSync, Meditor, AppEvolve, APIMigrator, A3, LASE, AndroEvolve}, or configuration files~\cite{openrewrite}. 
Such approaches commonly rely on knowledge about which APIs correspond across versions and on concrete examples of how API usages should be adapted. 
More recently, LLM-based techniques have been explored for updating client code that depends on Android API~\cite{GUPPY}, SQLAlchemy~\cite{SQLAlchemy}, and JUnit~\cite{GoogleCodeMigration}. 
These techniques similarly benefit from API replacement knowledge and representative update examples. 
API update mappings, their applicable version transitions, and client update instances therefore are essential for developing and evaluating automated library update techniques. 

Recent library evolution datasets, as summarized in Table~\ref{tab: existing_datasets}, focus on tasks such as version-aware code completion~\cite{Nizar2024-Arxiv, Kuhar2025-NAACL, Wang2025-ICSE} and code updates~\cite{Wu2024-Arxiv, CODEMENV}. 
However, they remain limited in the API update information they provide, the number of third-party libraries they cover, and the inclusion of update instances from real client dependency updates. 

First, existing datasets do not simultaneously provide API update mappings, version transitions, and client update instances. 
GitChameleon~\cite{Nizar2024-Arxiv} and LibEvolutionEval~\cite{Kuhar2025-NAACL} focus on code completion for specific library versions without providing explicit API update mappings or version transitions. 
Wang \textit{et al.}~\cite{Wang2025-ICSE} provide API update mappings but do not associate them with version transitions. 
VersiCode Editing~\cite{Wu2024-Arxiv} and CodeMEnv~\cite{CODEMENV} capture API update mappings across version transitions, but their code examples are not from real client dependency updates. 
Consequently, none of these datasets connects API update mappings and version transitions with client update instances. 

Second, existing datasets cover a small number of third-party libraries. 
Most focus exclusively on Python~\cite{Nizar2024-Arxiv, Kuhar2025-NAACL, Wang2025-ICSE, Wu2024-Arxiv}, covering 8 to 16 third-party libraries. 
CodeMEnv~\cite{CODEMENV} supports both Java and Python, but covers only four third-party Python libraries and no third-party Java libraries. 
Real projects, however, commonly depend on dozens or hundreds of third-party libraries~\cite{Huang2022-EMSE, Alfadel2023}. 
Consequently, datasets constructed from a small collection of libraries may not capture the diversity of API updates across third-party libraries. 

Third, existing datasets do not obtain update instances from real client dependency updates. 
Their code examples may be handwritten~\cite{Nizar2024-Arxiv}, generated by LLMs~\cite{CODEMENV}, or derived from documentation and docstrings~\cite{Wu2024-Arxiv, Kuhar2025-NAACL}. 
Such examples are useful for controlled evaluation, but may not capture how independent clients adapt API usages during library updates~\cite{APIfix, A3}. 
Even datasets~\cite{Wang2025-ICSE, Kuhar2025-NAACL} that obtain code from GitHub do not systematically connect API call changes with dependency version transitions in which clients perform the updates. 

Taken together, existing library evolution datasets provide complementary forms of API update information, but none simultaneously connects API update mappings, version transitions, and client update instances across a large number of third-party libraries. 

\begin{table*}[t]
\setlength{\tabcolsep}{2pt}
\renewcommand{\arraystretch}{1.1}
\footnotesize
\centering
\caption{Comparison of representative library evolution datasets in terms of API update information, coverage, and code sources. Version Transition indicates whether the dataset explicitly associates an API update mapping with source and target library versions. Client Update Instances indicates whether the dataset contains API call changes observed in actual client dependency updates.} 
\begin{threeparttable}
\begin{tabular}{
    >{\raggedright\arraybackslash}m{0.19\linewidth}
    >{\raggedright\arraybackslash}m{0.11\linewidth}
    >{\centering\arraybackslash}m{0.12\linewidth}
    >{\centering\arraybackslash}m{0.10\linewidth}
    >{\centering\arraybackslash}m{0.13\linewidth}
    >{\centering\arraybackslash}m{0.13\linewidth}
    >{\raggedright\arraybackslash}m{0.14\linewidth}}
\toprule
\textbf{Dataset} &
\textbf{Language} &
\textbf{\# API Update Mappings} &
\textbf{Version Transition} &
\textbf{Client Update Instances} &
\textbf{\# Third-party Libraries} &
\textbf{Code Source} \\
\midrule

GitChameleon~\cite{Nizar2024-Arxiv}
& Python
& \xmark
& \xmark
& \xmark
& 11
& Handwritten \\

LibEvolutionEval~\cite{Kuhar2025-NAACL}
& Python
& \xmark
& \xmark
& \xmark
& 8
& GitHub \& documentation \\

Wang \textit{et al.}~\cite{Wang2025-ICSE}
& Python
& 145
& \xmark
& \xmark
& 8
& GitHub \\

VersiCode Editing~\cite{Wu2024-Arxiv}
& Python
& 83
& \cmark
& \xmark
& 16
& Docstrings \\

CodeMEnv~\cite{CODEMENV}
& Java/Python
& 0/136
& \cmark
& \xmark
& 0/4\tnote{*}
& LLM-generated \\

\midrule

\textbf{\toolname}
& \textbf{Java/Python}
& 18,900/4,456
& \cmark
& \cmark
& \textbf{2,557/999}
& \textbf{Client updates} \\

\bottomrule
\end{tabular}

\begin{tablenotes}
\footnotesize
\item[*] CodeMEnv additionally contains 7 Python and 8 Java built-in libraries.
\end{tablenotes}
\end{threeparttable}
\label{tab: existing_datasets}
\end{table*}

\subsection{Mining API Update Mappings and Client Update Instances}
The library-driven workflow for constructing ecosystem-scale library update datasets generally involves two steps. 
The first step is to identify API update mappings. 
One source is library documentation, including changelogs, release notes, and upgrade guides~\cite{10123507}. 
Such documentation can explicitly describe recommended replacements, but its availability and completeness vary substantially across projects~\cite{10.1145/3524610.3527919}, and it may not describe updates between nonadjacent library versions~\cite{Meditor}. 
Prior work has also inferred API correspondences using library developers' input~\cite{565039, 10.1145/1806799.1806832, 10.1145/1062455.1062512} or API definition information such as signatures and implementations~\cite{1401931, 1566154, 4359473, Aura, LibSync, Huang2021-ASE, APIfix}. 
Other approaches analyze API usage changes within the library itself~\cite{Schafer, SemDiff}. 
Such internal usages provide useful evidence for identifying API correspondences, but may not reflect how external clients use and adapt APIs during library updates. 
Scaling these mapping identification approaches to a large library ecosystem requires analyzing API evolution across numerous library releases. 
Given the millions of library releases on registries such as Maven Central~\cite{mvnrepository} and PyPI~\cite{pypi}, exhaustively considering version pairs can incur prohibitive analysis cost, while restricting analysis to selected libraries reduces the coverage of the resulting dataset. 

The second step is to obtain client update instances for the identified API update mappings. 
It requires locating projects that use the relevant library, perform an applicable version transition, and exercise the affected APIs among hundreds of millions of public repositories~\cite{githubOctoverseDeveloper}. 
Client API usage is also sparse: projects generally use only a small subset of the APIs exposed by their dependencies~\cite{Wang2020-ICSME, MLCatchUp, Wang2025-ICSE}. 
Consequently, mappings identified from library artifacts may have few or no observable client update instances~\cite{Zhang2020SANER, HARRAND2022111134, Wang2025-ICSE}. 

In summary, scaling the library-driven workflow requires identifying API update mappings across numerous library releases and locating their client update instances among large numbers of client projects. 
Consequently, constructing ecosystem-scale library update datasets that connect API update mappings, version transitions, and client update instances remains challenging. 

\subsection{World of Code}
World of Code (WoC)~\cite{Ma2019-MSR, Ma2021-EMSE} is a large-scale research infrastructure for analyzing version control data. 
It aggregates Git objects~\cite{gitscmObjects}, including commits, trees, and blobs\footnote{A blob stores the content of a particular file version.}, from public repositories hosted on platforms such as GitHub, Bitbucket, and GitLab. 
WoC organizes these objects into key--value databases that support efficient cross-referencing among commits, files, blobs, and projects, and provides APIs for querying these databases and retrieving raw object contents. 
This infrastructure enables dependency file changes and their corresponding source code changes to be queried at ecosystem scale without locally cloning hundreds of millions of repositories, making it suitable for mining client dependency updates. 
In this study, we use WoC V3, updated in May 2024, which contains more than 4.7 billion commits and 19.6 billion blobs across 234 million repositories~\cite{wocOverview}. 

\section{The \toolname Framework}\label{s: method}
This section first introduces the client-driven workflow adopted by \toolname and then describes the design and language-specific implementation of each phase. 

\subsection{Overview}
To automatically construct ecosystem-scale library update datasets that connect API update mappings, version transitions, and client update instances, we propose \toolname, a client-driven framework built on WoC. 
We instantiate \toolname for Java and Python for two reasons. 
First, they are among the most widely used programming languages. 
Second, they differ in language design (static vs. dynamic type, compiled vs. interpreted) and library packaging mechanism, allowing us to examine whether the client-driven framework can be instantiated across substantially different ecosystems. 

Unlike the library-driven dataset construction workflow discussed in Section~\ref{s: intro}, \toolname adopts a client-driven workflow that starts from dependency updates observed in client projects. 
A dependency update directly identifies the updated library and its old and new versions, while associated source code changes provide candidate adaptations for that version transition. 
From these changes, \toolname mines candidate update instances, validates them against the involved library versions, and derives API update mappings from the validated instances. 
Consequently, each retained mapping is associated with a version transition and grounded in at least one client update instance. 

\begin{figure*}
    \centering
    \includegraphics[width=\linewidth]{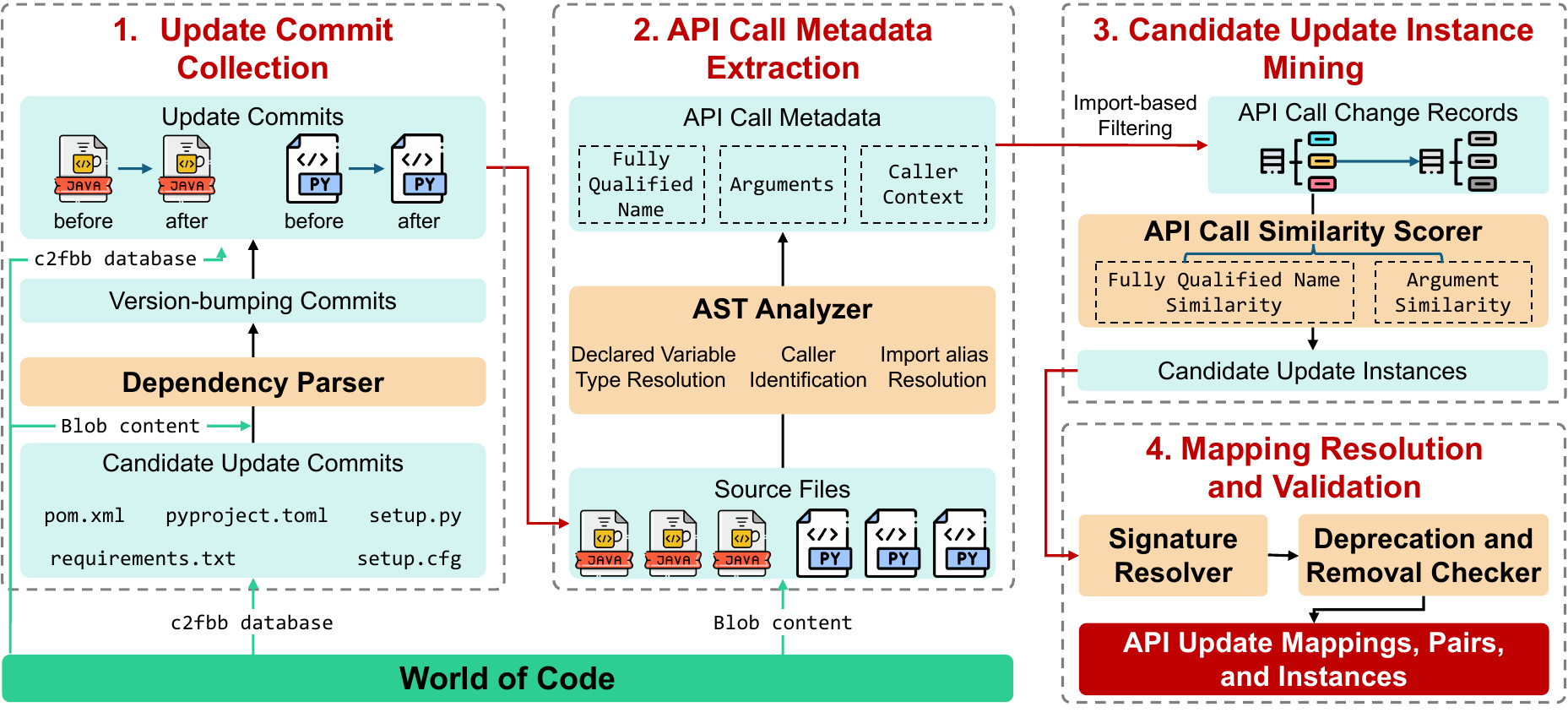}
    \caption{Overview of the \toolname framework. Light blue blocks represent data artifacts, light orange blocks represent components requiring language-dependent implementations, and the dark red block represents the final dataset.}
    \Description{Bridge collects dependency update commits, extracts API call metadata, mines candidate update instances with similarity-based matching, and resolves and validates API mappings against library artifacts. Light blue blocks represent data artifacts, light orange blocks represent components with language-dependent implementations, and the dark red block represents the final dataset of validated API update mappings and instances.}
    \label{fig: overview}
\end{figure*}

\toolname consists of four phases, as show in Figure~\ref{fig: overview}. 
First, \textit{Update Commit Collection} identifies commits in WoC that update dependency versions and also modify source files, which naturally links version transitions and client code changes. 
Second, \textit{API Call Metadata Extraction} parses source files before and after each dependency update and extracts API call metadata. 
Third, \textit{Candidate Update Instance Mining} identifies changed calls associated with the updated library and applies similarity-based matching to obtain plausible legacy--replacement API call pairs (i.e., candidate update instances). 
Finally, \textit{Mapping Resolution and Validation} resolves the involved API signatures for each candidate and validates the candidate using library-side evidence. 
From the validated instances, \toolname derives API update mappings and API update pairs.

\subsection{Phase 1: Update Commit Collection}\label{ss: phase-1}
\textbf{Design}. 
The client-driven workflow begins by identifying commits in which a client updates a library dependency and also modifies source code. 
Such commits provide two pieces of information needed by subsequent phases: library version transitions and associated client code changes from which candidate update instances can be mined. 
Consistent with prior work~\cite{Meditor, APIfix, PyMigTax}, \toolname assumes that a library update and its associated source code adaptations are typically performed within the same commit. 
The collection process progressively filters commits according to changes in dependency configuration files, dependency versions, and source files. 

\toolname first queries WoC for candidate update commits that modify dependency configuration files. 
WoC provides the \Code{c2fbb} database, which maps each commit to the paths of changed files and the SHA-1 hashes of their old and new blobs. 
Because programming language ecosystems commonly use recognizable dependency configuration files, such as \Code{pom.xml} for Maven projects, \toolname uses their filenames to retrieve such commits from \Code{c2fbb}. 
A modification to a dependency configuration file does not necessarily represent a library update because a commit may instead add, remove, or reorganize dependencies~\cite{MigrationAdvisor, SALM}. 
\toolname therefore parses the old and new configuration blobs, extracts dependency names and version constraints, and retains commits that change the version of at least one dependency as version-bumping commits. 
Following prior work~\cite{Meditor, Kuhar2025-NAACL}, we consider only pinned version constraints (e.g., \Code{==5.2}), which allow the old and new dependency versions to be identified precisely. 
Range-based constraints (e.g., \Code{>=3.1.0}) and unconstrained dependencies are excluded because their exact resolved versions cannot be determined from the configuration file alone~\cite{10.1145/3460319.3464797, 10.1145/3576037, Wang2020-ICSE, Huang2020-FSE}. 
Finally, it queries \Code{c2fbb} again and retains version-bumping commits that also modify source files, yielding the \emph{update commits} processed in subsequent phases. 

\textbf{Implementation for Java}. 
In the Java ecosystem, \Code{pom.xml} is the primary Maven dependency configuration file and has been widely used to extract dependencies from Java projects~\cite{Ochoa2022, MigrationAdvisor, 10.1145/3597926.3598147}. 
We therefore use changes to \Code{pom.xml} to identify candidate update commits. 
We implement a specialized XML parser following the Maven POM specification~\cite{pomxml} to extract dependency names and version constraints from relevant elements, such as \Code{<dependency>}. 
As shown in Table~\ref{tab: phase1_statistics}, we identify 31.9 million candidate update commits associated with 64.0 million unique \Code{pom.xml} blobs, of which 6.1 million (19.2\%) are version-bumping commits. 
Among these version-bumping commits, 17.2\% also modify Java source files, yielding 1.0 million update commits and 27.3 million unique Java source blobs. 
Collectively, these commits involve 50,450 distinct Java libraries and 630,405 library versions. 

\begin{table*}
\footnotesize
\centering
\caption{Statistics for data mined in Phase 1. CU, VB, and CFG are short for candidate update commits, version-bumping commits, and configuration blobs, respectively.}
\label{tab: phase1_statistics}
\renewcommand{\arraystretch}{1.1}
\setlength{\tabcolsep}{3pt}
\begin{tabular}{llrrrrrrr}
    \toprule
    \multirow{2}{*}{\textbf{Language}} & \multirow{2}{*}{\textbf{Configuration File}} & \multicolumn{3}{c}{\textbf{\# Commits}} & \multicolumn{2}{c}{\textbf{\# Blobs}} & \multirow{2}{*}{\textbf{\# Libraries}} & \multirow{2}{*}{\textbf{\# Versions}} \\
    & & \# CU & \# VB & \# Update & \# CFG & \# Source & & \\
    \midrule
    Java & pom.xml & 31,852,062 & 6,103,952 & 1,049,834 & 63,959,850 & 27,252,177 & 50,450 & 630,405 \\
    \midrule
    \multirow{5}{*}{Python} & requirements.txt & 14,677,782 & 7,719,071 & 672,536 & 10,189,499 & 6,745,851 & 22,889 & 210,326 \\
    & setup.py & 7,622,076 & 238,418 & 78,275 & 5,525,611 & 1,394,187 & 6,337 & 58,544 \\
    & pyproject.toml & 2,725,506 & 64,965 & 16,586 & 2,045,404 & 266,739 & 1,830 & 10,898 \\
    & setup.cfg & 1,174,085 & 29,832 & 7,886 & 131,717 & 719,294 & 1,442 & 7,857 \\
    \cmidrule[\lightrulewidth]{2-9}
    & \textbf{Total} & 25,002,699 & 7,993,598 & 753,902 & 18,479,562 & 7,794,556 & 25,397 & 236,292 \\
    \bottomrule
\end{tabular}
\end{table*}

\textbf{Implementation for Python}. 
Python projects use multiple configuration files to declare dependencies. 
We consider four common files: \Code{requirements.txt}, \Code{setup.py}, \Code{pyproject.toml}, and \Code{setup.cfg}. 
Unlike the executable \Code{setup.py}, the other three files represent dependency declarations in structured formats: plain text, TOML, and INI, respectively. 
We therefore implement specialized parsers following their specifications~\cite{requirementstxt, pyprojecttoml, setupcfg}. 
Parsing \Code{setup.py} is more challenging because dependencies can be programmatically declared through the \Code{install\_requires} and \Code{extras\_require} parameters of the \Code{setup()} function provided by \Code{setuptools}~\cite{pypaKeywordsSetuptools}. 
The \Code{install\_requires} parameter specifies a list of dependency strings (e.g., \Code{["BazSpam==1.1"]}), whereas \Code{extras\_require} maps optional feature names to dependency lists (e.g., \Code{\{"PDF": ["ReportLab>=1.2"]\}}). 
To statically recover these declarations from executable \Code{setup.py} files, we implement the lightweight data-flow analysis formalized in Algorithm~\ref{alg: setup.py_parser}. 

\begin{algorithm}
\caption{\texttt{setup.py} Parser}
\label{alg: setup.py_parser}
\DontPrintSemicolon
\Input{The source code of a \texttt{setup.py} blob: $source$}
\Output{Dependency names and version constraints: $\mathcal{D}$}

\SetKwFunction{DFG}{build\_DFG}
\SetKwFunction{import}{is\_import\_node}
\SetKwFunction{alias}{parse\_alias}
\SetKwFunction{assignment}{is\_assignment\_node}
\SetKwFunction{addEdge}{resolve\_edges}
\SetKwFunction{setupCall}{is\_setup\_call}
\SetKwFunction{exist}{exists}
\SetKwFunction{dummy}{create\_dummy\_edges}
\Fn{\DFG{AST}}{
    aliases $\gets \emptyset$, DFG $\gets \emptyset$\;
    \For(\tcp*[f]{top-down traversal}){node $n \in AST$}{
      \If{\import{n}}{
        aliases $\gets$ aliases $\cup$ \alias{n}\;
      }
      \If{\assignment{n}}{
        \addEdge{n.lhs, n.rhs, DFG}\;
      }
      \If{\setupCall{n, aliases}}{
        \For{$kw \in n.kwargs$}{
            \If(\tcp*[f]{setup(install\_requires=...) case}){\exist{kw.key}}{
                \addEdge{kw.key, kw.value, DFG}\;
            }\Else(\tcp*[f]{setup(**kwargs) case}){
                \dummy{``install\_requires'', kw[``install\_requires''], DFG}\;
                \dummy{``extras\_require'', kw[``extras\_require''], DFG}\;
            }
        }
      }
    }
    \Return{DFG}\;
}

\SetKwFunction{parse}{parse}
\SetKwFunction{traverse}{traverse}
\Fn{\parse{source}}{
    Parse $source$ into an abstract syntax tree $AST$\;
    DFG $\gets$ \DFG{AST}\;
    $\mathcal{D} \gets$ \traverse{``install\_requires'', DFG} $\cup$ \traverse{``extras\_require'', DFG}\;
    \Return{$\mathcal{D}$}\;
}
\end{algorithm}

The parser first constructs an abstract syntax tree (AST) from the source code and then performs a top-down traversal to build a data-flow graph (DFG) (lines 17--18). 
The DFG represents data dependencies among program elements. 
Its nodes may represent variables (e.g., \Code{a}), subscripts (e.g., \Code{a["b"]} and \Code{l[0]}), and constant values. 
A variable node is represented by its identifier, whereas a subscript node is represented by its target and index (e.g., \Code{(a, "b")} and \Code{(l, 0)}). 
To track values stored in lists and dictionaries, the parser further decomposes their elements and items into individual DFG nodes. 
The DFG is then constructed using the following rules: 
\begin{itemize}[leftmargin=*, topsep=0pt]
    \item For an import statement (e.g., \Code{from setuptools import setup}), the parser records the mapping from the imported alias (e.g., \Code{setup}) to its fully qualified name (e.g., \Code{setuptools.setup}) (lines 4--5), enabling subsequent identification of calls to the \Code{setup} API. 
    \item For an assignment, the parser creates DFG nodes for the left-hand side (LHS) and right-hand side (RHS) respectively and adds directed edges from the LHS node to the RHS node (lines 6--7). 
    \item For a call to the \Code{setuptools.setup} or \Code{distutils.core.setup} API, the parser similarly creates edges from keyword parameters to their actual arguments (lines 8--11). 
    To handle keyword argument unpacking (e.g., \Code{setup(**kwargs)}), it creates dummy nodes for \Code{install\_requires} and \Code{extras\_require} and connects them to corresponding items in the unpacked argument (lines 13--14). 
\end{itemize}
\noindent Finally, the parser traverses the DFG from the \Code{install\_requires} and \Code{extras\_require} nodes, collects reachable constant strings, and extracts dependency names and version constraints from the collected strings (line 19). 

In total, we identify 25.0 million candidate update commits associated with 18.5 million unique Python configuration blobs. 
Parsing their dependency declarations identifies 8.0 million (32.0\%) version-bumping commits, of which 9.4\% also modify Python source files, yielding 753,902 update commits and 7.8 million unique Python source blobs. 
Collectively, these update commits involve 25,397 distinct Python libraries and 236,292 library versions. 
Among the four configuration files, \Code{requirements.txt} contributes the largest number of update commits, followed by \Code{setup.py}, consistent with their widespread use for dependency management in Python projects. 

\subsection{Phase 2: API Call Metadata Extraction}\label{ss: phase-2}
\textbf{Design}. 
Given the update commits and associated old and new source blobs collected in Phase 1, \toolname extracts structured metadata for API calls in each blob. 
This metadata transforms raw client code changes into call-level information for similarity-based matching in Phase 3 and API signature resolution in Phase 4. 
Specifically, \toolname extracts three types of metadata for each API call: 
\begin{itemize}[leftmargin=*]
    \item \textbf{Fully Qualified Name (FQN)}. 
    The FQN identifies the called API and constitutes a fundamental part of an API signature. 
    It has been widely used to compare Java APIs~\cite{LibSync, Huang2021-ASE}. 
    \item \textbf{Actual Arguments}. 
    Actual arguments provide call-site information for assessing whether an API call removed from the old blob and one added to the new blob may represent a replacement. 
    In addition to argument expressions, \toolname extracts language-specific information useful for matching and signature resolution, including statically resolved argument types for Java and keyword names for Python. 
    \item \textbf{Caller Context}. 
    We represent the caller context of an API call as the sequence of enclosing declaration nodes from the AST root to the call node. 
    Phase 3 compares removed and added API calls within the same caller context, based on the intuition that API replacements associated with a library update are typically performed in place. 
    This constraint also reduces the number of call pairs that need to be compared. 
\end{itemize}

Precisely resolving all API calls could require building each client project and resolving its dependencies, which is computationally prohibitive at the scale of the update commits collected in Phase 1. 
\toolname therefore performs lightweight static analysis on individual source blobs without constructing project build environments. 
The analysis includes import alias resolution, caller identification, and declared variable type resolution. 
We implement the analysis for both languages using \Code{tree-sitter}~\cite{tree-sitter}, a high-performance parser generator that provides concrete syntax trees and query support for locating relevant constructs such as API calls, import statements, and variable declarations. 
The language-specific implementations are described below. 

\begin{algorithm}
\caption{Java API Call Metadata Extractor}
\label{alg: java-api-call-metadata-extractor}
\DontPrintSemicolon
\Input{The source code of a Java source blob: $source$}
\Output{API call metadata: $\mathcal{M}$}
\SetKwFunction{caller}{identify\_caller}
\SetKwFunction{dec}{is\_declaration\_node}
\SetKwFunction{add}{add}
\Fn{\caller{node}}{
    $\mathcal{C} \gets \emptyset$, cur\_node $\gets$ node\;
    \While(\tcp*[f]{bottom-up traversal to extract caller context}){cur\_node $\neq$ ROOT}{
        \If{\dec{cur\_node}}{
            $\mathcal{C}$.\add{cur\_node}\;
        }
        cur\_node $\gets$ cur\_node.parent\;
    }
    \Return{$\mathcal{C}$}\;
}
\SetKwFunction{objType}{extract\_variable\_declared\_types}
\SetKwFunction{varDec}{query\_variable\_declaration\_nodes}
\Fn{\objType{AST}}{
    $\mathcal{T} \gets \emptyset$\;
    \For(\tcp*[f]{find all variable declaration nodes}){node $d \in$ \varDec{AST}}{
        caller $\gets$ \caller{d}\;
        $\mathcal{T}$.\add{d.name, d.type, caller}\;
    }
    \Return{$\mathcal{T}$}\;
}
\SetKwFunction{parse}{extract\_api\_call\_metadata}
\SetKwFunction{import}{parse\_import\_nodes}
\SetKwFunction{apiCall}{query\_api\_call\_nodes}
\SetKwFunction{resolveFQN}{resolve\_fully\_qualified\_name}
\SetKwFunction{resolveArgs}{resolve\_argument\_types}
\Fn{\parse{source}}{
    $\mathcal{M} \gets \emptyset$\;
    Parse $source$ into an abstract syntax tree $AST$\;
    import\_name\_mappings $\gets$ \import{AST}\;
    variable\_types $\gets$ \objType{AST}\;
    \For(\tcp*[f]{extract metadata for each API call}){node $c \in$ \apiCall{AST}}{
        caller $\gets$ \caller{c}\;
        fqn $\gets$ \resolveFQN{c.name, caller, variable\_types, import\_name\_mappings}\;
        arguments $\gets$ \resolveArgs{c.arguments, caller, variable\_types}\;
        $\mathcal{M}$.\add{fqn, arguments, caller}\;
    }
    \Return{$\mathcal{M}$}\;
}
\end{algorithm}

\begin{figure*}
\centering
    \subfigure[Java API call metadata extraction]{\includegraphics[width=\linewidth]{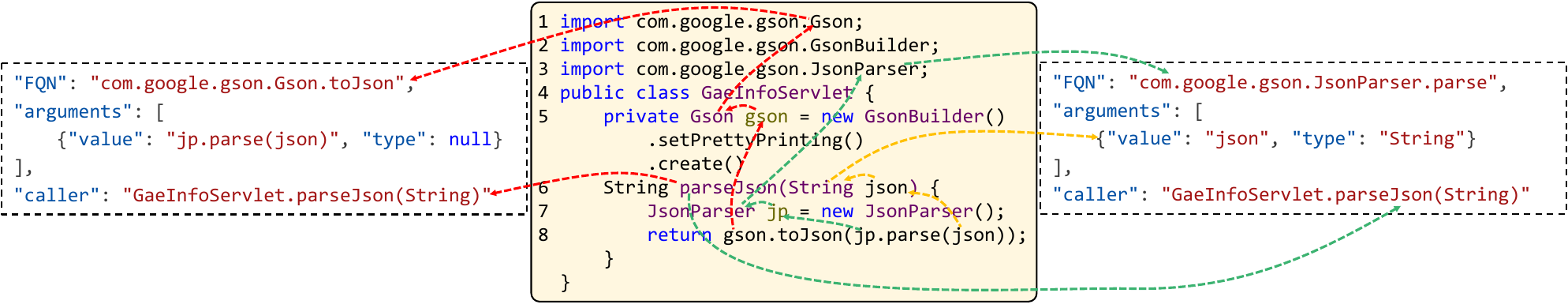}\label{fig: java-api-call-example}}
    \subfigure[Python API call metadata extraction]{\includegraphics[width=\linewidth]{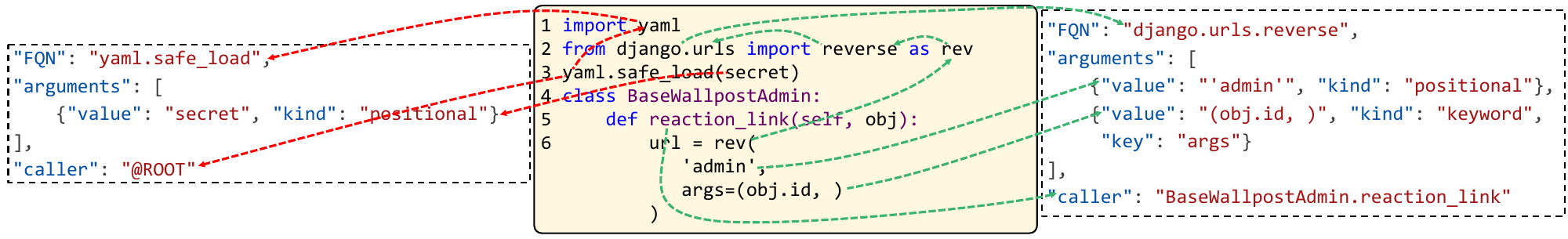}\label{fig: python-api-call-example}}
    \caption{Examples of API call metadata extraction for Java and Python.}
    \Description{Java and Python code examples illustrate how Bridge resolves fully qualified names, actual arguments, argument types, and caller contexts from API calls.}
\label{fig: api-call-example}
\end{figure*}

\textbf{Implementation for Java}. 
The Java implementation performs import resolution, declared variable type resolution, caller identification, and API call metadata extraction, as summarized in Algorithm~\ref{alg: java-api-call-metadata-extractor} and illustrated in Figure~\ref{fig: java-api-call-example}. 
\begin{enumerate}[leftmargin=*]
    \item \textit{AST Generation}. 
    The extractor first parses each Java source blob into an AST using \Code{tree-sitter} (line 16). 
    \item \textit{Import Resolution}. 
    It locates import statement nodes using a \Code{tree-sitter} query and records mappings from imported class names to their FQNs (line 17). 
    For example, for the first import statement in Figure~\ref{fig: java-api-call-example}, it records \Code{Gson} $\mapsto$ \Code{com.google.gson.Gson}. 
    These mappings are subsequently used to resolve FQNs for API calls (line 21). 
    \item \textit{Variable Type Extraction}. 
    The extractor collects explicitly declared variable types throughout the source file (line 18). 
    A \Code{tree-sitter} query identifies relevant declaration nodes, including fields, local variables, formal parameters, try-with-resources variables, and variables declared in enhanced \Code{for} statements (line 11). 
    For each declaration, the extractor records the variable name, declared type, and caller context (line 12). 
    The caller context is identified through a bottom-up traversal from the declaration node to the AST root that records enclosing declaration nodes, such as classes, methods, and interfaces (lines 1--7). 
    For example, the extractor records (\Code{json}, \Code{String}, \Code{GaeInfoServlet.parseJson(String)}) for the formal parameter in the sixth statement in Figure~\ref{fig: java-api-call-example}. 
    \item \textit{API Call Metadata Extraction}. 
    Finally, the extractor locates API call nodes using another \Code{tree-sitter} query and extracts metadata for each call (line 19). 
    It first identifies the call's caller context (line 20). 
    For a call with a receiver object, it attempts to resolve the API FQN using the receiver's declared type (line 21). 
    Specifically, it searches the recorded variable type information in the current caller context and its enclosing contexts for the receiver's declared type. It then restores the call's API FQN by searching the declared type against the import mappings. 
    The extractor also records argument expressions and attempts to resolve their declared types from the recorded variable type information (line 22). 
    Figure~\ref{fig: java-api-call-example} shows the metadata extracted for the two API calls in the eighth statement. 
    \toolname cannot statically determine the type of the argument in \Code{gson.toJson(jp.parse(json))} because doing so would require resolving the return type of another API call. 
    \toolname intentionally omits such interprocedural or dependency-aware type resolution to keep metadata extraction scalable across millions of source blobs. 
\end{enumerate}

\textbf{Implementation for Python}. 
The Python implementation focuses on import alias resolution and caller identification. 
Because Python is dynamically typed, reliably resolving variable types through lightweight static analysis is difficult and time-consuming. 
We therefore do not perform variable type inference and instead resolve API FQNs when they can be determined from import and alias information. 
Figure~\ref{fig: python-api-call-example} illustrates the extraction process. 
\begin{enumerate}[leftmargin=*]
    \item \textit{AST Generation}. 
    The extractor first parses each Python source blob into an AST. 
    \item \textit{Alias Resolution}. 
    It locates import statement nodes using a \Code{tree-sitter} query and records mappings from imported names and aliases to their FQNs. 
    For example, for the second statement in Figure~\ref{fig: python-api-call-example}, it records \Code{rev} $\mapsto$ \Code{django.urls.reverse}. 
    These mappings are used to resolve FQNs for imported API calls. 
    \item \textit{API Call Metadata Extraction}. 
    Finally, the extractor locates API call nodes using another \Code{tree-sitter} query and, for each call, attempts to resolve its FQN, records its argument expressions, and identifies its caller context using the same bottom-up procedure as the Java implementation. 
    For each argument, it further distinguishes positional and keyword arguments and records the keyword name when present. 
    This syntactic information is useful for comparing Python API calls because keyword names explicitly identify argument roles. 
    For example, the call \Code{rev(...)} in the sixth statement contains a positional argument \Code{'admin'} and a keyword argument \Code{args=(obj.id, )}. 
    Figure~\ref{fig: python-api-call-example} shows the metadata extracted for the two API calls. 
\end{enumerate}

\subsection{Phase 3: Candidate Update Instance Mining}\label{ss: phase-3}
\textbf{Design}. 
Using the API call metadata extracted in Phase 2, Phase 3 mines candidate update instances associated with each dependency version transition identified in Phase 1. 
The key intuition is that, when a client adapts its code to an updated dependency, an API call removed from the old code may be replaced by a related API call added to the new code within the same caller context. 
\toolname therefore first identifies changed API calls belonging to the updated library and then matches removed and added calls based on their FQN and argument similarities. 
At this stage, these similarities provide only client-side replacement signals and the mined candidate update instances are further validated using library-side evidence in Phase 4. 

\textit{API Call Change Detection}. 
The first challenge is to determine which API calls belong to the updated library. 
A Library and its API FQNs are not directly linked, because a published library may expose one or more importable prefixes that differ from its name. 
\toolname therefore constructs a mapping from each updated library to its importable top-level prefixes. 
Specifically, it downloads the latest available artifact of each updated library and extracts top-level import prefixes from the artifact. 
For Java, these prefixes are derived from class file paths in JARs, whereas for Python they are derived from importable Python module paths in wheels. 
Using the latest artifact avoids downloading every historical release and assumes that a library's top-level import prefixes remain sufficiently stable across releases for identifying calls belonging to that library. 

\toolname associates each changed source file with dependency updates occurred in its nearest configuration file. 
It then uses the import prefix mapping to identify API calls belonging to each updated library and groups calls by library and caller context. 
Within each group, \toolname compares the calls extracted from the old and new blobs and retains calls whose FQNs occur exclusively on one side. 
Calls appearing only in the old blob and new blob form $C_o$ and $C_n$, respectively. 
Calls with the same FQN on both sides are excluded even when their arguments differ. 
This conservative design focuses the mining process on replacements between different APIs and avoids interpreting regular argument changes, overloaded Java calls, or changes enabled by Python's flexible argument passing as API replacements. 
Each API call change record is represented as $(c, f, l, v_o, v_n, ctx, b_o, b_n, C_o, C_n)$, where $c$ denotes the update commit, $f$ the source file path, $l$ the updated library, $v_o$ and $v_n$ the old and new library versions, $ctx$ the caller context, $b_o$ and $b_n$ the old and new blob hashes, and $C_o$ and $C_n$ the removed and added API calls, respectively. 

\textit{Threshold-Based Bipartite Matching}. 
An API call change record may contain multiple removed and added calls within the same caller context. 
\toolname models all possible correspondences between $C_o$ and $C_n$ as a weighted bipartite graph, where each edge $(c_o, c_n)$ represents a possible replacement and is assigned a language-specific similarity score computed from the calls' FQN and actual arguments. 
As formalized in Algorithm~\ref{alg:matching}, \toolname computes the similarity of every removed--added call pair and discards edges whose scores do not exceed a language-specific threshold $\theta$. 
It then sorts the remaining edges by descending similarity and greedily selects edges subject to a one-to-one constraint. 
The selected edges form candidate update instances and are passed to Phase 4 for library-side resolution and validation. 

\begin{algorithm}
\caption{Threshold-based Bipartite Matching}
\label{alg:matching}
\DontPrintSemicolon
\Input{An API call change record $r=(c, f, l, v_o, v_n, ctx, b_o, b_n, C_o, C_n)$, similarity function $S$, threshold $\theta$}
\Output{Candidate update instances $\mathcal{I}$}
$E \gets \emptyset$\;
\For(\tcp*[f]{Pairwise API call similarity calculation}){$i \gets 1$ \KwTo $|C_o|$}{
    \For{$j \gets 1$ \KwTo $|C_n|$}{
        $s \gets S(C_o[i], C_n[j])$\;
        \If(\tcp*[f]{Keep edges above the similarity threshold}){$s > \theta$}{
            $E \gets E \cup \{(i, j, s)\}$\;
        }
    }
}
Sort $E$ by descending similarity $s$, then ascending old API call index $i$\;
$\mathcal{I} \gets \emptyset$, $U_o \gets \emptyset$, $U_n \gets \emptyset$\;
\For{$(i,j,s) \in E$}{
    \If(\tcp*[f]{Enforce one-to-one matching}){$i \notin U_o$ \textbf{and} $j \notin U_n$}{
        $\mathcal{I} \gets \mathcal{I} \cup \{(l, v_o, C_o[i], v_n, C_n[j])\}$\;
        $U_o \gets U_o \cup \{i\}$\;
        $U_n \gets U_n \cup \{j\}$\;
    }
}
\Return{$\mathcal{I}$}\;
\end{algorithm}

\textbf{Implementation for Java}. 
Table~\ref{tab: api_change_statistics} summarizes the API call changes detected before bipartite matching. 
For Java, \toolname identifies 342,018 API call change records from 42,885 update commits (4.1\% of those collected in Phase 1) and 222,243 unique code blobs. 
These records involve 5,352 libraries, 31,084 library versions, and 138,396 unique API FQNs. 

To measure the similarity between two Java API calls, \toolname uses three signals: Class Name Similarity ($CS$), Method Name Similarity ($MS$), and Argument Similarity ($AS_j$). 
For $CS$ and $MS$, identifiers are tokenized according to CamelCase and snake\_case conventions, and token overlap is measured using a Jaccard-style score. 
To reduce high similarity between lexically related but semantically contrasting names, \toolname applies conservative rules that set name similarity to zero for patterns such as different leading action verbs (e.g., \Code{get} versus \Code{set}), incompatible primitive type tokens, different logging levels, and mismatched negation markers. 
For common factory-style method names such as \Code{of}, \Code{from}, and \Code{getInstance}, which carry little replacement information by themselves, \toolname uses the class name as the method name signal. 
For $AS_j$, \toolname compares the ordered lists of actual argument expressions using the Levenshtein ratio. 
The resulting Java API Call Similarity ($JACS$) combines the three signals as follows: 
\begin{equation}
    JACS(c_o, c_n) = w_{CS} \times CS(c_o, c_n) + w_{MS} \times MS(c_o, c_n) + w_{AS_j} \times AS_j(c_o, c_n)
\end{equation}

\begin{table}
\centering
\caption{Statistics of API call changes detected before bipartite matching.}
\label{tab: api_change_statistics}
\renewcommand{\arraystretch}{1.1}
\begin{tabular}{lrrrrrr}
    \toprule
    \textbf{Language} & \textbf{\# Records} & \textbf{\# Commits} & \textbf{\# Blobs} & \textbf{\# Libraries} & \textbf{\# Versions} & \textbf{\# API FQNs} \\
    \midrule
    Java & 342,018 & 42,885 & 222,243 & 5,352 & 31,084 & 138,396 \\
    Python & 172,999 & 31,379 & 125,861 & 2,641 & 15,540 & 36,804 \\
    \bottomrule
\end{tabular}
\end{table}

\textbf{Implementation for Python}. 
As shown in Table~\ref{tab: api_change_statistics}, \toolname identifies 172,999 Python API call change records from 31,379 update commits (4.2\% of those collected in Phase 1) and 125,861 unique code blobs. 
These records involve 2,641 libraries, 15,540 library versions, and 36,804 unique API FQNs. 
Because Phase 2 does not infer receiver or argument types, \toolname measures Python API call similarity using FQN Similarity ($FS$) and Argument Similarity ($AS_p$). 

For $FS$, \toolname primarily compares the final component of each FQN, corresponding to the function or method name. 
Inspection of API mappings in prior library evolution datasets~\cite{Wang2025-ICSE, Wu2024-Arxiv, CODEMENV} suggests that replacement APIs may express related functionality using different naming conventions or abbreviations. 
Accordingly, \toolname first normalizes names by removing underscores and converting characters to lowercase, such that \Code{arg\_max} and \Code{argmax} receive full similarity. 
It also treats a compact name as equivalent to an expanded name when the compact form consists of the initials of the expanded tokens, as in \Code{mse} and \Code{mean\_squared\_error}. 
Otherwise, names are tokenized according to snake\_case and CamelCase conventions, and token overlap is measured using a Jaccard-style score. 

For $AS_p$, \toolname compares arguments according to Python's argument passing mechanisms. 
Specifically, Python allows four types of arguments~\cite{pythonMoreControl}: positional arguments (e.g., \Code{arg}), variadic positional arguments (e.g., \Code{*args}), keyword arguments (e.g., \Code{mode="fast"}), and variadic keyword arguments (e.g., \Code{**kwargs}). 
Positional arguments and variadic positional arguments are compared as ordered sequences using a weighted edit distance, where insertion and deletion have a cost of one and substitution has a cost of two. 
Their contribution is computed as the total length of the two sequences minus their edit distance. 
Keyword arguments are compared using both keys and values: matching keys contribute to the score, while matching values provide additional evidence. 
Variadic keyword arguments are treated as sets of unpacked argument expressions, and their contribution is the size of their exact intersection. 
The contributions from all the four types of arguments are summed and normalized by the total number of arguments in the two calls to obtain $AS_p$. 
For example, consider \Code{f(x, y, mode="fast")} and \Code{g(z, y, mode="safe")}. 
Their positional argument sequences, \Code{[x, y]} and \Code{[z, y]}, have an edit distance of 2 and therefore contribute $|[x,y]| + |[z,y]| - 2 = 2$. 
The keyword argument \Code{mode} contributes one additional unit because its key matches, whereas its values provide no additional contribution because \Code{"fast"} and \Code{"safe"} differ. 
The three matched argument units are normalized by the six actual arguments in the two calls, giving $AS_p=(2+1)/6=0.5$. 

The resulting Python API Call Similarity ($PACS$) combines FQN and argument similarities as follows: 
\begin{equation}
PACS(c_o, c_n) = w_{FS} \times FS(c_o, c_n) + w_{AS_p} \times AS_p(c_o, c_n)
\end{equation}

\subsection{Phase 4: Mapping Resolution and Validation}\label{ss: phase-4}
\textbf{Design}. 
Phase 3 generates candidate update instances by similarity-based matching of client-side API call changes, but these similarity signals alone are insufficient to validate the candidates. 
Phase 4 therefore resolves the API signatures involved in each candidate and validates the candidate using library-side API evolution evidence. 

Candidate update instances mined in Phase 3 may contain false positives for two main reasons. 
First, Phase 3 associates API calls with updated libraries through import prefixes, but different libraries may expose the same prefixes, causing calls to be incorrectly attributed to the updated library. 
Second, similar API calls does not necessarily indicate an API replacement. 
Calls with similar FQNs and arguments may provide different functionality, while, particularly in Python, calls with different FQNs may resolve to the same underlying API through re-exports. 

\textit{API Signature Resolution}. 
To ensure that the paired calls belong to the updated library and are available in the involved versions, \toolname resolves their API signatures against the library artifacts corresponding to the version transition. 
As shown in Algorithm~\ref{alg:mappingvalidation}, for each candidate $(l,v_o,c_o,v_n,c_n)$, \toolname resolves the call $c_o$ against version $v_o$ of library $l$ and the call $c_n$ against version $v_n$. 
If either call cannot be resolved to an API exposed by its respective library artifact, the candidate is discarded. 
This version-specific resolution can filter candidates incorrectly attributed to the updated library. 

\textit{Removal and Deprecation Validation}. 
Successful signature resolution alone is insufficient to establish that a candidate represents an update instance. 
\toolname therefore additionally requires evidence that the legacy API is removed from or deprecated in the new library version. 
To obtain this evidence, \toolname attempts to resolve the call $c_o$ against the new library version. 
If $c_o$ can no longer be resolved, \toolname treats this absence as evidence that the legacy API has been removed. 
If $c_o$ remains resolvable but the resolved API carries deprecation evidence, \toolname treats the candidate as supported by deprecation. 
Otherwise, the candidate is discarded. 
This conservative criterion favors precision over exhaustive coverage and intentionally excludes replacements for which the legacy API remains available without detectable deprecation evidence. 

For each retained candidate, the resolved legacy and replacement API signatures $a_o$ and $a_n$ form the API update mapping $(l,a_o,a_n)$. 
Associating this mapping with the version transition identified in Phase 1 yields the API update pair $(l,v_o,a_o,v_n,a_n)$, while the client-side call change forms the validated client update instance. 
Phase 4 therefore connects API update mappings, version transitions, and client update instances. 
Each language-specific implementation provides two capabilities for this validation: API signature resolution and deprecation checking. 

\begin{algorithm}
\caption{Mapping Resolution and Validation}
\label{alg:mappingvalidation}
\DontPrintSemicolon
\Input{Candidate update instances $\mathcal{I}$}
\Output{Validated update instances $\mathcal{V}$, mappings $\mathcal{M}$, and update pairs $\mathcal{P}$}
$\mathcal{V}, \mathcal{M}, \mathcal{P} \gets \emptyset$\;
\SetKwFunction{resolve}{resolve\_API\_signature}
\SetKwFunction{dep}{is\_deprecated}
\For{$r=(l, v_o, c_o, v_n, c_n) \in \mathcal{I}$}{
    $a_o \gets$ \resolve{$l$, $v_o$, $c_o$}\;
    \If(\tcp*[f]{The old call does not resolve to an API in the old library version}){$a_o=\emptyset$}{
        continue\;
    }
    $a_n \gets$ \resolve{$l$, $v_n$, $c_n$}\;
    \If(\tcp*[f]{The new call does not resolve to an API in the new library version}){$a_n=\emptyset$}{
        continue\;
    }
    $a_o^n \gets$ \resolve{$l$, $v_n$, $c_o$}\;
    \If(\tcp*[f]{The legacy API is removed or deprecated in the new library version}){$a_o^n = \emptyset$ or \dep{$a_o^n$}}{
        $\mathcal{V} \gets \mathcal{V} \cup \{r\}$\;
        $\mathcal{M} \gets \mathcal{M} \cup \{(l, a_o, a_n)\}$\;
        $\mathcal{P} \gets \mathcal{P} \cup \{(l, v_o, a_o, v_n, a_n)\}$\;
    }
}
\Return{$\mathcal{V}, \mathcal{M}, \mathcal{P}$}\;
\end{algorithm}

\textbf{Implementation for Java}. 
\textit{API Signature Resolution}. 
As defined in Section~\ref{ss: terminology}, a Java API signature consists of an FQN and its formal parameter types~\cite{oracleDefiningMethods}. 
Because Java supports method overloading, an FQN may resolve to multiple API signatures. 
Given an API call and a library version, \toolname downloads the sources JAR for that version and resolves the call against method definitions in its Java and Scala source files. 
The resolver first locates the source file exposing the call's FQN, parses it with \Code{tree-sitter}, and searches for method definitions with the same FQN. 
It also follows class inheritance when the called method may be inherited from a superclass or implemented interface defined within the library. 
Candidate definitions are filtered based on whether their numbers of formal parameters are compatible with the actual arguments. 
When multiple signatures remain, \toolname uses the argument information extracted in Phase 2 to identify the most plausible signature. 
Specifically, it compares the declared types of the actual arguments with the formal parameter types of each candidate signature and selects the signature with the highest type compatibility. 

\textit{Deprecation Checking}. 
For a resolved API, \toolname checks for deprecation evidence following common deprecation practices in Java projects. 
Specifically, it recognizes the \Code{@Deprecated} annotation and the \Code{@deprecated} Javadoc tag. 
Deprecation is checked at both the method and enclosing-class levels because an API may become deprecated when its declaring class is deprecated. 
A Java candidate is retained only when the old call resolves to an API in the old version, the new call resolves to an API in the new version, and the legacy API is removed from or identified as deprecated in the new version. 

\textbf{Implementation for Python}. 
\textit{API Signature Resolution}. 
For Python, \toolname uses the FQN as the API signature because Python does not use formal parameter types to distinguish overloaded methods. 
Given an API call and a library version, \toolname downloads the wheel for that version and resolves the call against APIs exposed in its Python files. 
The resolver first identifies the Python file exposing the API by performing longest-prefix matching between the call's FQN and Python file paths in the wheel. 
It then parses the matched file and determines whether the remaining FQN components resolve to a function, class, or class member defined in the file. 
If an FQN component is imported from another Python file, the resolver follows the import and repeats the resolution procedure on the target file. 

\textit{Deprecation Checking}. 
For a resolved API, \toolname searches its definition for deprecation evidence. 
Prior work~\cite{Wang2020-FSE} shows that Python libraries deprecate APIs through diverse mechanisms, including decorators, warning messages, and documentation text. 
Following this observation, \toolname examines decorators of the API's definition and searches its docstrings and string literals for deprecation cues, including ``deprecated'', ``removed'', ``legacy'', ``renamed'', and ``moved''. 
A Python candidate is retained only when the old call resolves to an API in the old version, the new call resolves to an API in the new version, and the legacy API is removed from or identified as deprecated in the new version. 

\section{Evaluation}\label{s: evaluation}
We evaluate \toolname with respect to its accuracy of mining client update instances, the impact of its key design choices, the characteristics of the mined dataset, and the utility enabled by the dataset. 
Specifically, we investigate the following research questions (RQs): 
\begin{itemize}[leftmargin=*, topsep=0pt]
    \item \textbf{RQ1:} \textbf{How accurately does \toolname mine client update instances from API call changes?} \\
    We evaluate the precision, recall, and F1 score of the client update instances produced by Phases 3 and 4 against a manually annotated ground truth dataset. 
    \item \textbf{RQ2:} \textbf{How do similarity-based matching and library-side validation affect the effectiveness of \toolname?} \\
    We analyze the sensitivity of \toolname to alternative similarity weights and matching thresholds and quantify the contribution of Phase 4 through an ablation study. 
    \item \textbf{RQ3:} \textbf{What are the scale and distributional characteristics of the dataset mined by \toolname?} \\
    We characterize the mined API update mappings, API update pairs, and client update instances across libraries and versions, with particular attention to how mappings and instances are distributed. 
    \item \textbf{RQ4:} \textbf{What does the mined dataset reveal about LLM-based replacement API recommendation?} \\
    As one example application of the dataset, we evaluate four large language models (LLMs) on replacement API recommendation, a key step in automated library updates, and investigate how recommendation performance relates to the frequency of API update mappings in client updates. 
\end{itemize}

\subsection{Ground Truth Dataset}
RQ1 and RQ2 aim to evaluate the effectiveness of \toolname in mining client update instances, particularly its similarity-based candidate matching in Phase 3 and library-side validation in Phase 4. 
We therefore construct a manually annotated ground truth dataset from the API call change records identified in Phase 3. 
Starting from these records allows us to evaluate the two key designs that determine which API call change pairs are retained as client update instances: Phase 3 matches removed and added calls to generate candidate update instances, while Phase 4 resolves their API signatures and validates the candidates using library-side API evolution evidence. 

\textbf{Sampling}. 
Naively sampling from all API call change records could cause frequently updated libraries and repeatedly changed APIs to dominate the ground truth dataset. 
We therefore adopt a two-stage sampling strategy to increase diversity across both libraries and API calls. 
For each language, we first sample 100 libraries with probabilities proportional to their numbers of update commits, giving libraries with more observed update activity a greater probability of selection while maintaining coverage across libraries. 
For each selected library, we then randomly sample API call change records and retain a record only if its old-side API FQN has not been previously selected for that library. 
We continue sampling until at least ten distinct old-side API FQNs are covered or no additional records remain for the library. 
This procedure reduces repeated observations of frequently changed APIs and increases the diversity of API call changes represented in the ground truth. 
Accordingly, the ground truth is designed to evaluate \toolname across diverse libraries and API calls rather than to estimate population-weighted accuracy over all API call change records. 
The resulting ground truth dataset contains 638 Java and 609 Python API call change records. 

\textbf{Annotation}. 
The first two authors independently inspect each sampled record and label all possible one-to-one pairs between its old-side and new-side API calls. 
A pair is labeled as a genuine client update instance only when external evidence supports that the new-side API replaces the old-side API under the observed library version transition. 
The annotators seek evidence from official API documentation, release notes and upgrade guides, and the library's source repository; when necessary, they additionally consult Stack Overflow discussions. 
For each positive instance, they record the supporting evidence together with the legacy and replacement API signatures. 
We compute inter-rater agreement over all 2,283 Java and 1,988 Python API call pairs, with genuine client update instances labeled as positive and all other pairs as negative. 
The two annotators achieve a Cohen's $\kappa$ of 0.92, indicating almost perfect agreement. 
They subsequently discuss all disagreements and determine the final annotations by consensus. 
The resulting ground truth contains 344 Java and 242 Python client update instances. 

\subsection{Mining Effectiveness (RQ1)}
\subsubsection{Approach}
We evaluate the effectiveness of \toolname in mining client update instances from API call changes by running \toolname on the ground truth dataset and comparing the mined instances with the manual annotations. 
For Java, we assign equal weights of $1/3$ to $CS$, $MS$, and $AS_j$ and set the similarity threshold to 0.45. 
For Python, we assign equal weights of 0.5 to $FS$ and $AS_p$ and set the threshold to 0.35. 
RQ2 examines the sensitivity of these parameter choices and the contribution of library-side validation. 
A mined instance is considered a true positive only when its resolved legacy and replacement API signatures match those of an annotated client update instance. 

\subsubsection{Results}
\textbf{Overall effectiveness}. 
Table~\ref{tab:rq1-effectiveness} reports the performance of \toolname on the ground truth dataset. 
For Java, \toolname correctly identifies 305 of the 344 annotated instances, with 28 false positives and 39 false negatives, achieving 91.6\% precision, 88.7\% recall, and a 90.1\% F1 score. 
For Python, \toolname correctly identifies 155 of the 242 annotated instances, with 17 false positives and 87 false negatives, achieving 90.1\% precision, 64.0\% recall, and a 74.9\% F1 score. 
Thus, \toolname achieves precision above 90\% for both languages, consistent with our objective of favoring precision when automatically constructing a reusable dataset, while achieving substantially lower recall for Python than for Java. 

\begin{table*}
\centering
\caption{Performance of \toolname on the ground truth dataset. GT, TP, FP, and FN denote the numbers of annotated instances, true positives, false positives, and false negatives, respectively. Precision, recall, and F1 score are reported as percentages.}
\setlength{\tabcolsep}{4pt}
\renewcommand{\arraystretch}{1.1}
\begin{tabular}{lrrrrrrrrr}
\toprule
\textbf{Language} & \textbf{Records} & \textbf{GT} & \textbf{Predicted} & \textbf{TP} & \textbf{FP} & \textbf{FN} & \textbf{Precision} & \textbf{Recall} & \textbf{F1 score} \\
\midrule
Java   & 638 & 344 & 333 & 305 & 28 & 39 & 91.6 & 88.7 & 90.1 \\
Python & 609 & 242 & 172 & 155 & 17 & 87 & 90.1 & 64.0 & 74.9 \\
\bottomrule
\end{tabular}
\label{tab:rq1-effectiveness}
\end{table*}

\textbf{Error Analysis}. 
To understand the mining errors, we manually inspect every false positive and false negative and classify their primary causes. 

\textbf{False Positive Analysis}. 
Phase 4 verifies whether the calls in a candidate can be resolved to APIs in the involved library versions and whether the resolved legacy API is removed from or identified as deprecated in the new version. 
These conditions establish the availability and evolution status of the involved APIs, but do not necessarily establish a replacement relation between them. 
Consequently, an incorrect candidate produced by similarity-based matching in Phase 3 may still be retained through either the deprecation path, where the legacy API remains available but is identified as deprecated, or the removal path, where the legacy API can no longer be resolved in the new version. 
Of the 28 Java false positives, 4 (14.3\%) are retained through the deprecation path and 24 (85.7\%) through the removal path. 
In contrast, 13 (76.5\%) of the 17 Python false positives are retained through the deprecation path, while only 4 (23.5\%) are retained through the removal path. 
The prevalence of the deprecation path among Python false positives mainly stems from the permissive deprecation checking required to accommodate diverse Python deprecation practices. 
Specifically, \toolname examines decorators, docstrings, and string literals within the resolved API definition for deprecation cues. 
While this strategy captures diverse forms of deprecation evidence, unrelated text containing such cues may be incorrectly interpreted as deprecation evidence, allowing an incorrect candidate to survive validation. 

\textbf{False Negative Analysis}. 
False negatives arise from three main sources corresponding to similarity-based matching and library-side validation. 
First, an annotated instance may not exceed the similarity threshold and is therefore not selected as a candidate in Phase 3. 
Second, Phase 4 may fail to resolve one or both calls in a candidate to APIs in the involved library versions. 
Third, both calls may be resolved successfully, but the resolved legacy API remains available in the new version without deprecation evidence recognized by \toolname. 

For Java, 16 (41.0\%) of the 39 false negatives have similarity scores that do not exceed the selected threshold of 0.45 and are therefore not selected as candidates in Phase 3. 
Another 22 (56.4\%) result from failures to resolve one or both calls in Phase 4: 12 involve APIs inherited from classes defined outside the library, four involve methods generated by annotation processors, and six involve unsupported source constructs or unexpected source JAR layouts. 
The remaining false negative results from deprecation evidence not recognized by \toolname: the legacy API remains resolvable in the new version, while its deprecation is documented only in an issue comment in the library's source repository\footnote{\url{https://github.com/jpmml/jpmml-evaluator/issues/54\#issuecomment-313649349}}. 

For Python, 22 (25.3\%) of the 87 false negatives have similarity scores that do not exceed the selected threshold of 0.35 and are therefore not selected as candidates in Phase 3. 
Another 51 (58.6\%) result from failures to resolve one or both calls in Phase 4: 31 involve an unavailable wheel for at least one version, ten involve APIs exposed only at runtime, seven involve APIs defined in an external library, and three involve APIs defined in a binary module. 
The remaining 14 (16.1\%) result from deprecation evidence not recognized by \toolname because the involved libraries use deprecation mechanisms outside those supported by our checker. 
For example, PyYAML communicates the deprecation of certain uses of \Code{yaml.load} through runtime warnings issued by a dedicated helper function outside the API definition\footnote{\url{https://github.com/yaml/pyyaml/blob/0cedb2a0697b2bc49e4f3841b8d4590b6b15657e/lib/yaml/__init__.py\#L109}}. 
The definition-level checker therefore cannot capture this evidence. 
These results indicate that the lower Python recall primarily stems from API resolution and deprecation-checking limitations rather than similarity-based matching alone. 

\begin{resultbox}
\textbf{Answer to RQ1:} 
\toolname achieves precision, recall, and F1 scores of 91.6\%, 88.7\%, and 90.1\% for Java, and 90.1\%, 64.0\%, and 74.9\% for Python, respectively. 
The lower Python recall mainly results from limitations in API signature resolution and deprecation checking. 
Overall, precision above 90\% for both languages demonstrates that \toolname can mine client update instances from API call changes with high precision. 
\end{resultbox}

\subsection{Design Analysis (RQ2)}
\subsubsection{Approach}
We analyze two key design choices that determine which API call changes are retained as client update instances: the similarity-based matching used to generate candidates in Phase 3 and the library-side validation performed in Phase 4. 

\textit{Weight Sensitivity}. 
We enumerate positive weight configurations in increments of 0.1 that sum to one, producing 36 configurations for Java and nine for Python; we additionally evaluate the equal Java weights of $1/3$ used in RQ1. 
For each configuration, we conduct grouped 5-fold cross-validation, keeping all possible call pairs from the same API call change record in the same fold. 
In each round, we select the threshold between 0 and 1, in increments of 0.05, that maximizes the F1 score on the other four folds and then measure precision, recall, and F1 score on the held-out fold. 
We report the mean and standard deviation across the five folds. 

\textit{Threshold Sensitivity}. 
We fix the equal weights used in RQ1 and vary the matching threshold from 0 to 0.95 in increments of 0.05. 
At each threshold, we calculate the precision, recall, and F1 score. 
This analysis quantifies the precision--recall tradeoff of similarity-based candidate matching and motivates the selected thresholds of 0.45 for Java and 0.35 for Python. 

\textit{Phase 4 Ablation}. 
Phase 4 resolves the calls in each candidate against the involved library versions and retains only candidates for which the resolved legacy API is removed from or identified as deprecated in the new version. 
While this library-side validation filters false candidates produced by similarity-based matching, failures in API signature resolution or deprecation checking may also discard genuine client update instances. 
To quantify this tradeoff, we repeat the threshold analysis without Phase 4 and treat every candidate generated by Phase 3 as a client update instance. 
We compare the resulting precision, recall, and F1 score with those of the complete framework across thresholds and report the comparison at the thresholds selected in RQ1. 

\subsubsection{Results}
\begin{figure*}
\centering
\includegraphics[width=0.49\textwidth]{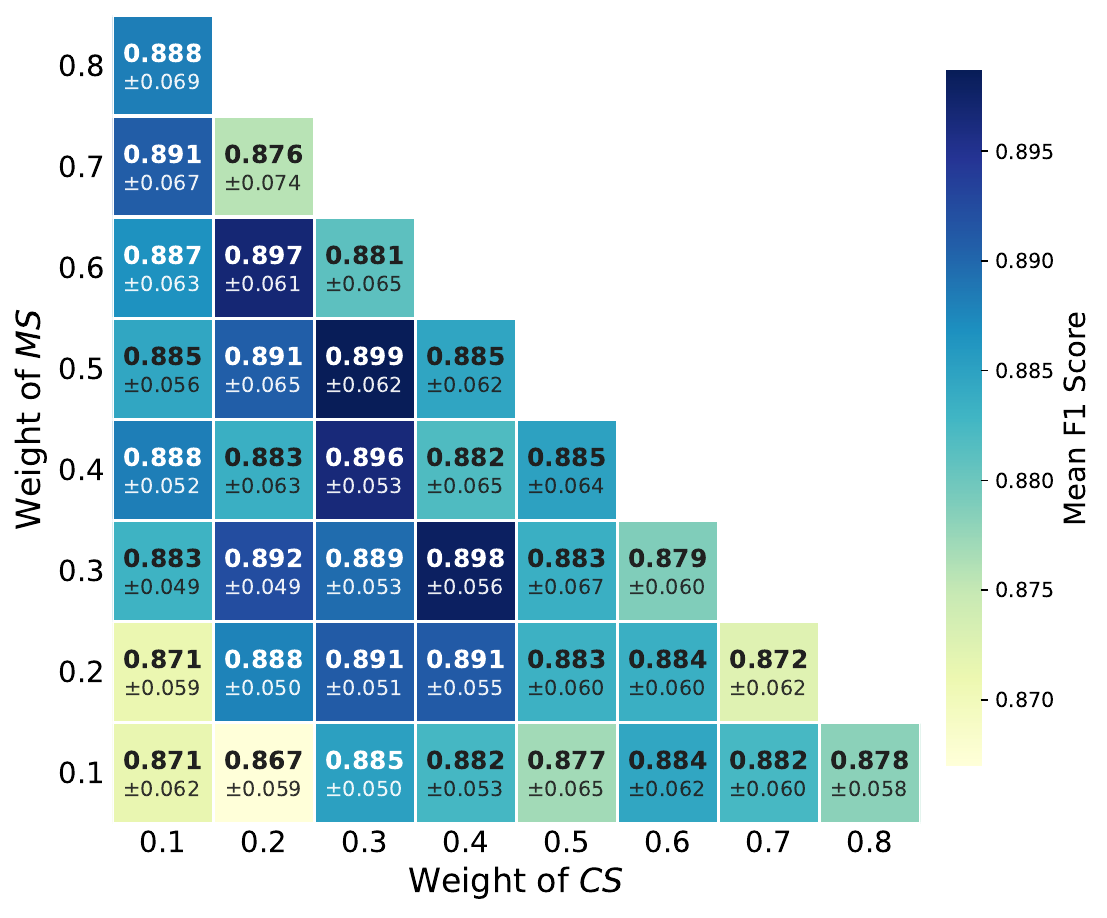}
\hfill
\includegraphics[width=0.49\textwidth]{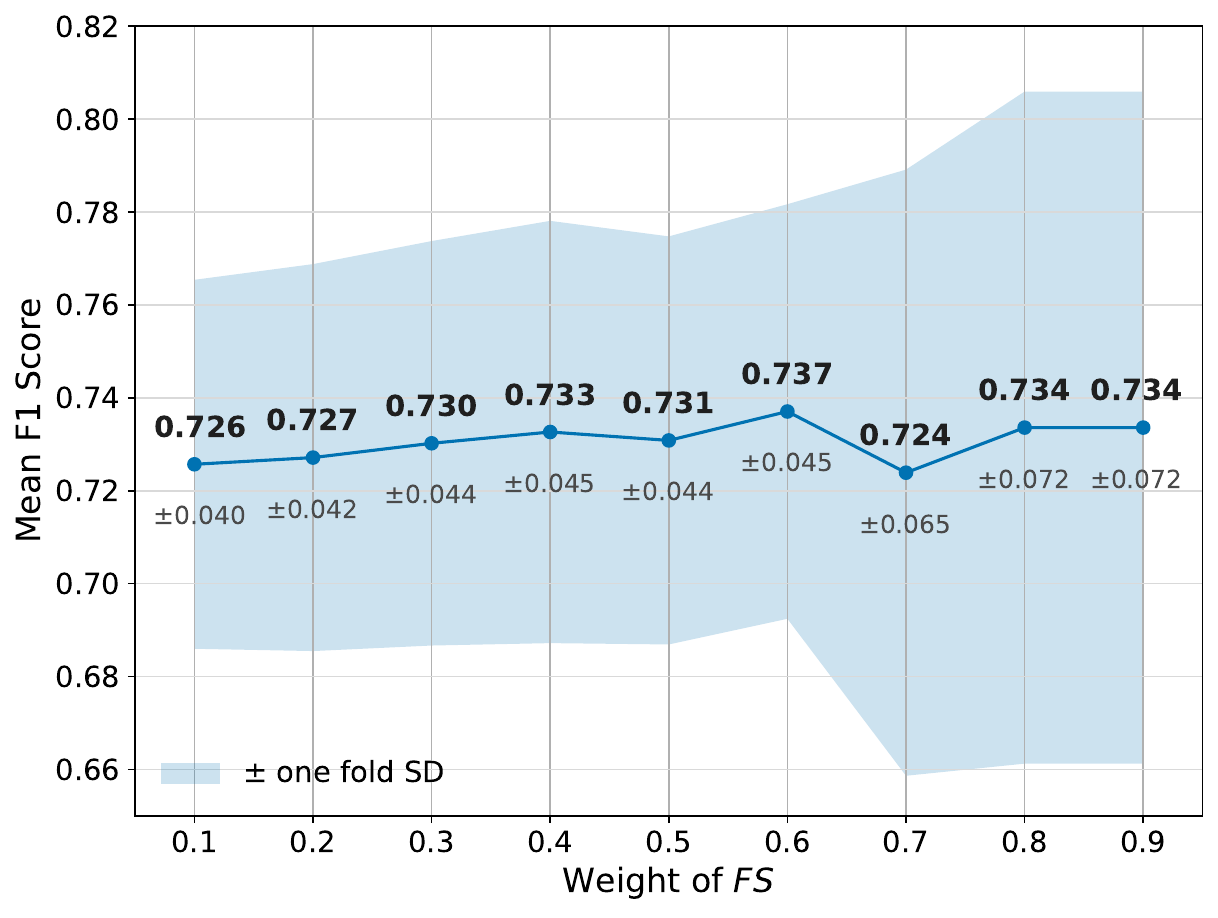}
\caption{Weight sensitivity for Java (left) and Python (right). Each value is the mean F1 score across five held-out folds, with its standard deviation. For Java, the argument similarity weight is $1-w_{CS}-w_{MS}$; the equal configuration is reported in the text because it does not lie on the 0.1 grid. For Python, the argument similarity weight is $1-w_{FS}$.}
\Description{The Java heatmap and Python line plot show only small differences in mean F1 across weight configurations.}
\label{fig:rq2-weight-sensitivity}
\end{figure*}

\textbf{Weight Sensitivity}. 
Figure~\ref{fig:rq2-weight-sensitivity} shows that the similarity weights have a limited effect on the F1 score. 
For Java, the mean F1 score ranges from 86.7\% to 89.9\% across the 37 configurations, a difference of 3.2 percentage points. 
The equal configuration obtains the highest mean F1 score of 89.9\% ($\mathit{SD}=5.5$ percentage points). 
For Python, the mean F1 score ranges from 72.4\% to 73.7\%, a difference of only 1.3 percentage points. 
The equal configuration reaches a mean F1 score of 73.1\% ($\mathit{SD}=4.4$ percentage points), only 0.6 percentage points below the highest value of 73.7\%. 
These results indicate that similarity-based matching is not highly sensitive to the weight configuration and support equal weighting as a simple and uniform choice across languages. 

\begin{figure*}
\centering
\includegraphics[width=\textwidth]{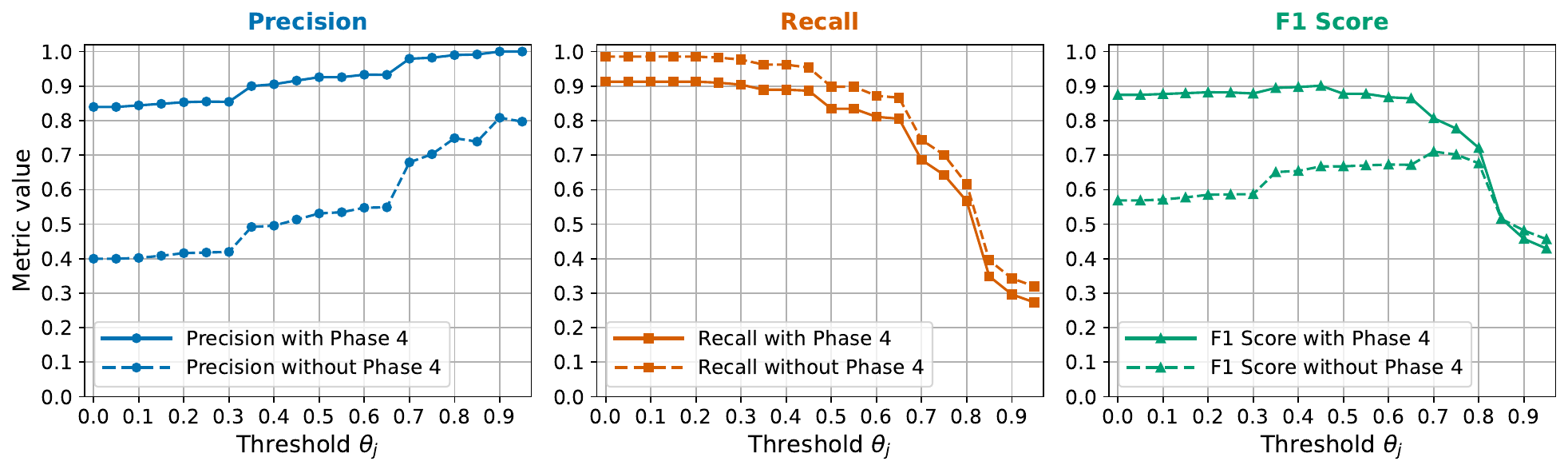}
\includegraphics[width=\textwidth]{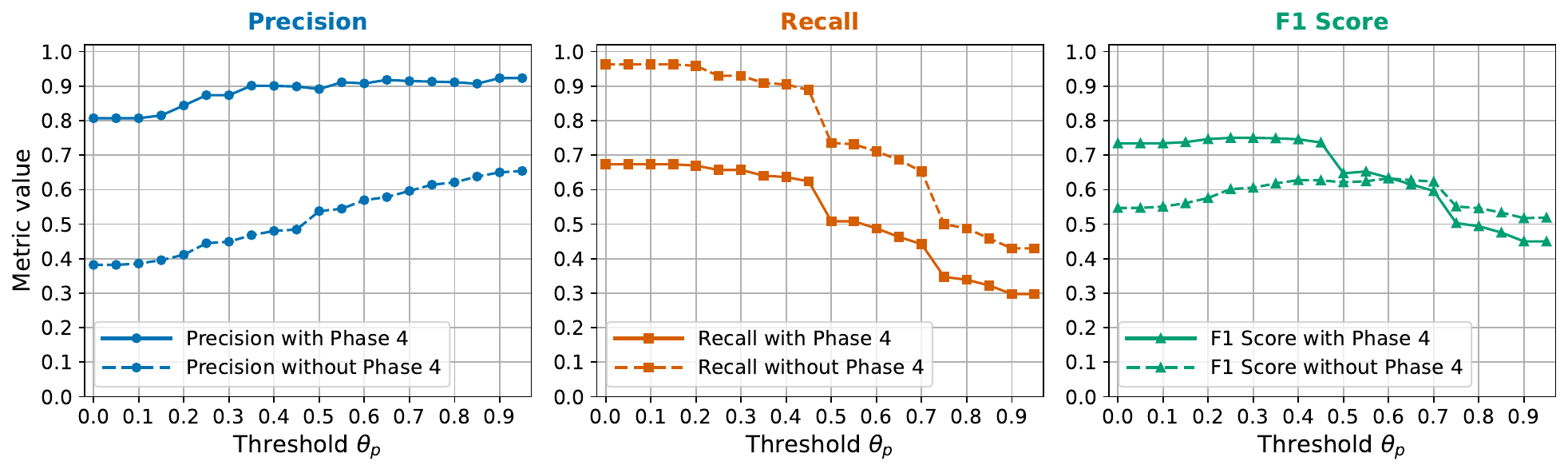}
\caption{Threshold sensitivity and Phase 4 ablation for Java (top) and Python (bottom). Solid lines show the complete framework, while dashed lines show the framework without Phase 4.}
\Description{Precision, recall, and F1 curves across thresholds show that higher thresholds favor precision over recall and that Phase 4 substantially improves precision for both languages.}
\label{fig:rq2-threshold-ablation}
\end{figure*}

\textbf{Threshold Sensitivity}. 
The solid curves in Figure~\ref{fig:rq2-threshold-ablation} show how the matching threshold controls the tradeoff between precision and recall. 
For Java, increasing the threshold from 0 to 0.45 raises precision from 84.0\% to 91.6\%, while decreasing recall from 91.3\% to 88.7\%. 
The selected threshold of 0.45 yields the highest F1 score of 90.1\%. 
For Python, increasing the threshold from 0 to 0.35 raises precision from 80.7\% to 90.1\%, while recall decreases from 67.4\% to 64.0\%. 
Although the F1 score is highest at 0.30, selecting 0.35 increases precision by 2.8 percentage points while reducing recall by 1.7 points and F1 by only 0.1 points. 
We therefore select 0.35 because precision is particularly important when automatically mined instances are released as a reusable dataset. 

\textbf{Ablation of Phase 4}. 
Table~\ref{tab:rq2-phase4-ablation} reports the ablation results at the selected thresholds, while the dashed curves in Figure~\ref{fig:rq2-threshold-ablation} show the effect across all thresholds. 
For Java, library-side validation increases precision from 51.3\% to 91.6\%, an improvement of 40.3 percentage points, while recall decreases from 95.3\% to 88.7\%. 
Consequently, the F1 score increases by 23.4 points, from 66.7\% to 90.1\%. 
For Python, library-side validation increases precision from 46.8\% to 90.1\%, an improvement of 43.3 percentage points, while recall decreases from 90.9\% to 64.0\%. 
The larger recall reduction for Python mainly stems from failures in API signature resolution and deprecation checking, as identified in the RQ1 false negative analysis. 
Nevertheless, the F1 score increases by 13.1 points, from 61.8\% to 74.9\%. 
These results demonstrate that library-side validation substantially improves the precision of similarity-based candidate matching for both languages, although this improvement comes with a larger recall reduction for Python. 

\begin{table*}
\small
\centering
\caption{Phase 4 ablation at the selected thresholds (0.45 for Java and 0.35 for Python).}
\renewcommand{\arraystretch}{1.1}
\begin{tabular}{lrrrrrr}
\toprule
\multirow{2}{*}{\textbf{Phase 4}} & \multicolumn{3}{c}{\textbf{Java}} & \multicolumn{3}{c}{\textbf{Python}} \\
% \cmidrule(lr){2-4}\cmidrule(lr){5-7}
& \textbf{Precision (\%)} & \textbf{Recall (\%)} & \textbf{F1 (\%)} & \textbf{Precision (\%)} & \textbf{Recall (\%)} & \textbf{F1 (\%)} \\
\midrule
With Phase 4    & 91.6 & 88.7 & 90.1 & 90.1 & 64.0 & 74.9 \\
Without Phase 4 & 51.3 & 95.3 & 66.7 & 46.8 & 90.9 & 61.8 \\
$\Delta$ & $\uparrow 40.3$ & $\downarrow 6.7$ & $\uparrow 23.4$ & $\uparrow 43.3$ & $\downarrow 26.9$ & $\uparrow 13.1$ \\
\bottomrule
\end{tabular}
\label{tab:rq2-phase4-ablation}
\end{table*}

\begin{resultbox}
\textbf{Answer to RQ2:} 
Similarity weights have limited impact on the F1 score and the matching threshold controls the precision--recall tradeoff of similarity-based candidate matching. 
At the selected thresholds, library-side validation improves precision by 40.3 percentage points for Java and 43.3 points for Python, demonstrating its critical role in achieving high-precision client update instance mining, although with a larger recall reduction for Python. 
\end{resultbox}

\subsection{Dataset Characteristics (RQ3)}
\subsubsection{Approach}
We apply the configured \toolname framework to the complete API call change record collection described in Section~\ref{ss: phase-3} and characterize the resulting library update dataset at multiple granularities. 
We first measure its scale in terms of libraries, releases, version transitions, API update mappings, API update pairs, client update instances, and distinct update commits. 
These units capture complementary aspects of the dataset: mappings represent API replacement relations, pairs associate mappings with version transitions, and instances capture the client-side API call changes in which the replacements are observed. 
We then report the numbers of client update instances retained through the removal and deprecation validation paths in Phase 4. 
Finally, we analyze the distributions of update commits, API update mappings, and client update instances across libraries, as well as the distribution of client update instances across mappings. 

\subsubsection{Results}
\textbf{Dataset Scale}. 
Table~\ref{tab:rq3-scale} summarizes the dataset mined from the complete collection. 
Phase 3 identifies 686,608 Java and 561,018 Python candidate update instances from 342,018 Java and 172,999 Python API call change records. 
Phase 4 discards 238,403 (34.7\%) Java and 85,926 (15.3\%) Python candidates because one or both calls cannot be resolved to APIs in the involved library versions, and another 66,544 (9.7\%) Java and 197,833 (35.3\%) Python candidates because the resolved legacy API is neither removed from nor identified as deprecated in the new version. 
After library-side validation, \toolname retains 381,661 Java client update instances from 17,266 update commits, representing 35,532 API update pairs and 18,900 API update mappings across 10,637 version transitions of 2,557 libraries. 
For Python, \toolname retains 277,259 client update instances from 7,850 update commits, representing 11,393 API update pairs and 4,456 API update mappings across 5,190 version transitions of 999 libraries. 
Among the retained Java instances, 269,736 (70.7\%) are retained through the deprecation path and 111,925 (29.3\%) through the removal path. 
The distribution is reversed for Python: 48,419 (17.5\%) are retained through the deprecation path and 228,840 (82.5\%) through the removal path. 

\begin{table*}
\footnotesize
\centering
\caption{Scale of the resulting dataset. Releases are distinct (library, version) combinations, version transitions are distinct (library, old version, new version) combinations.}
\renewcommand{\arraystretch}{1.1}
\setlength{\tabcolsep}{3pt}
\begin{tabular}{lrrrrrrr}
\toprule
\textbf{Language} & \textbf{\# Libraries} & \textbf{\# Releases} & \textbf{\# Version Transitions} & \textbf{\# Mappings} & \textbf{\# Pairs} & \textbf{\# Instances} & \textbf{\# Commits} \\
\midrule
Java   & 2,557 & 12,165 & 10,637 & 18,900 & 35,532 & 381,661 & 17,266 \\
Python & 999 & 4,908 & 5,190 & 4,456 & 11,393 & 277,259 & 7,850 \\
\bottomrule
\end{tabular}
\label{tab:rq3-scale}
\end{table*}

\begin{figure*}
\centering
\includegraphics[width=\textwidth]{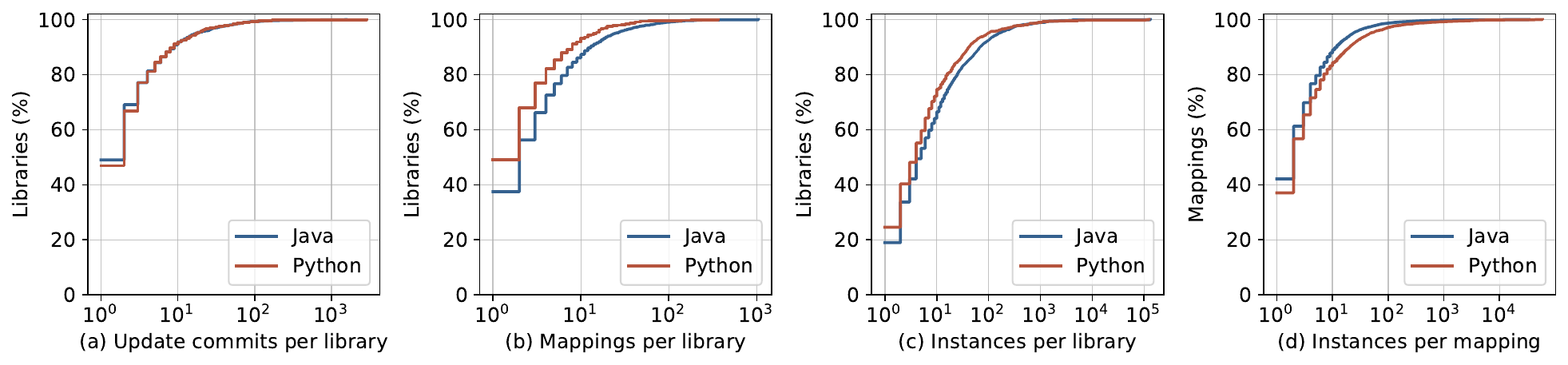}
\caption{Distributions of the resulting dataset. Each curve gives the percentage of libraries or mappings whose count is at most the corresponding x-axis value.}
\Description{Four empirical cumulative distributions compare Java and Python. Most libraries contain few update commits, mappings, and instances, and most mappings appear in few instances, while all distributions have long tails.}
\label{fig:rq3-dataset-characteristics}
\end{figure*}

\textbf{Distribution of Mappings and Instances}. 
Figure~\ref{fig:rq3-dataset-characteristics} shows the distributions of update commits, API update mappings, and client update instances across libraries, as well as client update instances across mappings. 
Across all four measures, both ecosystems exhibit highly skewed distributions, with most libraries or mappings accounting for relatively few observations and a small number accounting for substantially more. 

\textit{Commits per Library}. 
Java and Python exhibit similar distributions of update commits across libraries. 
The median library appears in two distinct update commits, with an interquartile range of one to three for both languages. 
Approximately half of the libraries appear in only one update commit (49.0\% for Java and 46.9\% for Python), and 91.9\% in both languages appear in at most ten update commits. 
At the other extreme, the Java library \Code{org.mockito:mockito-core} appears in 1,549 update commits and the Python library \Code{django} appears in 2,867 update commits, showing that a small number of libraries are represented by substantially more observed client updates. 

\textit{Mappings per Library}. 
The median library contains two API update mappings in both languages, with an interquartile range of one to five for Java and one to three for Python. 
Overall, 37.5\% of Java libraries and 49.1\% of Python libraries contain only one API update mapping, while 87.4\% and 93.2\%, respectively, contain at most ten mappings. 
The distributions nevertheless have long tails: the Java library \Code{org.elasticsearch:elasticsearch} contains 1,048 API update mappings, while the Python library \Code{tensorflow} contains 361 API update mappings. 

\textit{Instances per Library}. 
The median Java library appears in five client update instances, with an interquartile range of two to 17, whereas the median Python library appears in four, with an interquartile range of two to 11. 
Although 92.6\% of Java libraries and 95.1\% of Python libraries appears in at most 100 instances, the maxima reach 132,630 for the Java library \Code{junit:junit} and 124,226 for the Python library \Code{wagtail}. 
Thus, the hundreds of thousands of instances reported in Table~\ref{tab:rq3-scale} are highly unevenly distributed across libraries, with most libraries appearing in relatively few instances and a small number accounting for a large proportion of the dataset. 

\textit{Instances per Mapping}. 
The median API update mapping appears in two client update instances in both languages, with interquartile ranges of one to four for Java and one to six for Python. 
Overall, 42.1\% of Java mappings appear in only one client update instance and 89.0\% in at most ten instances. 
For Python, 37.1\% of mappings appear in only one instance and 84.4\% in at most ten instances. 
At the other extreme, the most frequent Java mapping, from \Code{junit.framework.Assert.assertEquals(int, int)} to \Code{org.junit.Assert.assertEquals(long, long)}, appears in 33,678 instances. 
The most frequent Python mapping, from \Code{django.core.urlresolvers.reverse} to \Code{django.urls.reverse}, appears in 58,780 instances. 
These results reveal a pronounced long-tail distribution of API update mappings: many mappings are observed in only one or a few client update instances, whereas a small number of recurrent mappings appear in thousands or even tens of thousands of instances. 

\begin{resultbox}
\textbf{Answer to RQ3:} 
\toolname mines 381,661 Java and 277,259 Python client update instances, representing 18,900 and 4,456 API update mappings across 2,557 and 999 libraries, respectively. 
The dataset exhibits highly skewed distributions across libraries and a pronounced long-tail distribution of API update mappings, with most mappings appearing in only a few client update instances and a small number appearing substantially more frequently. 
\end{resultbox}

\subsection{Dataset Utility (RQ4)}
\subsubsection{Approach}
The dataset mined by \toolname can support empirical studies of library evolution and the development and evaluation of automated library update techniques. 
As one example application, we use the dataset to evaluate LLMs on replacement API recommendation. 
Identifying the API that should replace a deprecated or removed API is a key step in automated library updates, yet the capability of current LLMs to perform this task remains unclear. 
The three connected units in the dataset support complementary parts of this evaluation: API update mappings provide reference replacements, API update pairs specify the applicable version transitions, and client update instances provide real-world call contexts. 
Moreover, the pronounced long-tail distribution identified in RQ3 enables us to examine how recommendation performance varies with the frequency of API update mappings in client update instances. 
We therefore investigate the overall recommendation accuracy, the effect of providing client call contexts, and the relationship between mapping frequency and recommendation performance. 

We conduct the evaluation under two input settings. 
The \textit{signature input} is constructed from an API update pair and provides the library, old version, new version, and legacy API signature. 
The \textit{context input} additionally provides focused client call context extracted from a client update instance. 
The context consists of relevant import statements and the legacy API call. 
For Java, the context additionally consists of variable declaration statements of the call's receiver object and arguments. 

\textbf{Evaluation Data}. 
A legacy API can occasionally be paired with multiple replacement APIs in the mined dataset because clients may replace the same API differently depending on its usage context. 
We therefore group client update instances by library, old version, new version, and legacy API, and retain the most frequent replacement only when it accounts for more than half of the instances in the group. 
This strict majority requirement reduces ambiguity among alternative replacements observed for the same legacy API and version transition. 
We then sample one API update pair for each distinct API update mapping and, for the context input, one associated client update instance. 
This procedure produces 18,166 Java and 4,360 Python examples. 
The two input settings use identical mappings, version transitions, and reference replacements. 

\textbf{Models and Prompts}. 
We evaluate GPT-5.4, DeepSeek-V4-Pro, DeepSeek-V4-Flash, and GLM-5.1. 
Listing~\ref{lst:rq4-prompt} presents the shared prompt template. 
For the signature input, \Code{<CONTEXT\_BLOCK>} is omitted; for the context input, it contains the focused client call context described above. 
The \Code{<SIGNATURE\_FORMAT>} requires a FQN with parameter types for Java and a FQN for Python. 
We use zero temperature and a maximum output length of 128 tokens, and disable optional thinking mode for endpoints that expose it. 

\begin{lstlisting}[style=prompt,caption={Prompt template for replacement API recommendation. Text in angle brackets denotes an example-specific placeholder.},label={lst:rq4-prompt}]
SYSTEM
You are an expert in library API evolution. Given a library version update and a legacy API, identify the API in the new library version that should replace the legacy API.
Return exactly one JSON object and no additional text: {"replacement_api": "<API signature>"}
The replacement must belong to the specified library and be available in the new version. If uncertain, return your single best recommendation.

USER
A project updates the following <LANGUAGE> library:
Library: <LIBRARY> 
Old version: <OLD_VERSION> 
New version: <NEW_VERSION> 
The following legacy API is deprecated or removed: <LEGACY_API>
<CONTEXT_BLOCK>
Recommend the replacement API in the new version and represent the replacement using <SIGNATURE_FORMAT>.
\end{lstlisting}

\textbf{Metrics}. 
We use exact-match accuracy as the primary metric. 
For Python, a prediction is correct when its FQN exactly matches the reference. 
For Java, full signature accuracy additionally requires the exact formal parameter types. 
We also report the FQN accuracy for Java, which ignores the parameter types, to distinguish replacement API identification from overload selection. 
To compare the two input settings, we use McNemar's exact test~\cite{mcnemar1947note} on their paired correctness outcomes for the same API update mapping and version transition, and apply Holm correction~\cite{holm1979simple} across the model and language comparisons. 

RQ3 reveals a pronounced long-tail distribution of API update mappings, with most mappings appearing in only a few client update instances and a small number appearing substantially more frequently. 
To examine how recommendation performance varies across this distribution, we analyze accuracy by \textit{mapping frequency}, defined as the number of validated client update instances associated with an API update mapping in the \toolname dataset. 
We divide mappings into five frequency groups: one, two to five, six to ten, 11 to 100, and more than 100 instances. 

\subsubsection{Results}
\textbf{Overall Recommendation Accuracy}. 
Table~\ref{tab:rq4-overall-accuracy} reports the recommendation accuracy across the four evaluated LLMs and two input settings. 
GPT-5.4 achieves the highest accuracy in every language and input setting. 
For Java, the full signature accuracy ranges from 21.2\% to 27.5\% under the signature input and from 20.6\% to 26.5\% under the context input. 
For Python, the FQN accuracy ranges from 30.0\% to 42.5\% and from 29.1\% to 44.4\%, respectively. 

\begin{table*}
\small
\centering
\caption{The recommendation accuracy across the four evaluated LLMs and two input settings. FQN and Full denote FQN accuracy and full signature accuracy, respectively.}
\renewcommand{\arraystretch}{1.1}
\begin{tabular}{lrrrrrr}
\toprule
\multirow{3}{*}{\textbf{Model}} & \multicolumn{4}{c}{\textbf{Java}} & \multicolumn{2}{c}{\textbf{Python}} \\
& \multicolumn{2}{c}{\textbf{Signature Input}} & \multicolumn{2}{c}{\textbf{Context Input}} & \textbf{Signature Input} & \textbf{Context Input} \\
& \textbf{FQN (\%)} & \textbf{Full (\%)} & \textbf{FQN (\%)} & \textbf{Full (\%)} & \textbf{FQN (\%)} & \textbf{FQN (\%)} \\
\midrule
GPT-5.4            & \textbf{37.1} & \textbf{27.5} & \textbf{36.1} & \textbf{26.5} & \textbf{42.5} & \textbf{44.4} \\
DeepSeek-V4-Pro    & 32.1 & 24.2 & 31.0 & 22.2 & 36.4 & 36.1 \\
DeepSeek-V4-Flash  & 25.7 & 21.2 & 25.8 & 20.6 & 30.0 & 29.1 \\
GLM-5.1            & 26.2 & 23.2 & 25.3 & 21.7 & 32.5 & 32.2 \\
\bottomrule
\end{tabular}
\label{tab:rq4-overall-accuracy}
\end{table*}

\begin{figure*}
\centering
\includegraphics[width=\textwidth]{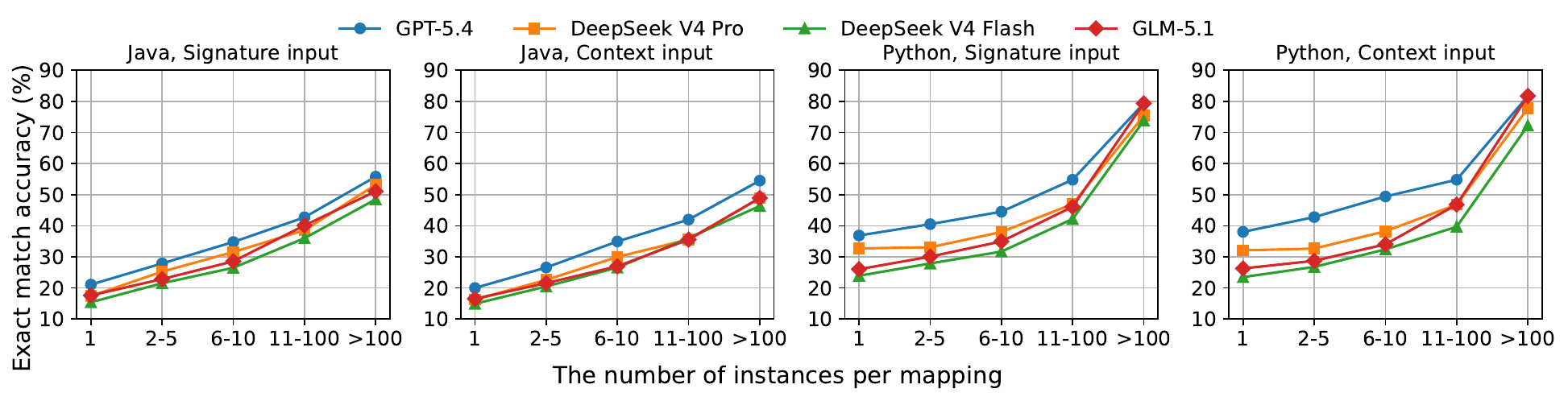}
\caption{Recommendation accuracy by mapping frequency. Java panels report full signature accuracy.}
\Description{Four line charts show replacement API recommendation accuracy for Java and Python under signature and context inputs. Accuracy generally increases with mapping frequency.}
\label{fig:rq4-mapping-frequency}
\end{figure*}

For Java, the FQN accuracy is 3.0 to 9.7 percentage points higher than full signature accuracy. 
Thus, some errors occur even when a model identifies the correct package, class, and method name because it predicts incorrect formal parameter types. 
Even when considering only FQNs, however, the highest accuracy reaches 37.1\% for Java and 44.4\% for Python. 
These results indicate that reliably identifying replacement APIs remains challenging for the evaluated LLMs. 

\textbf{Effect of Client Call Context}. 
Providing focused client call context does not consistently improve recommendation accuracy. 
For Java, the full signature accuracy decreases for all four models by 0.6 to 2.0 percentage points. 
The paired outcomes show that context changes predictions in both directions. 
For example, for GPT-5.4, context changes 959 incorrect predictions to correct ones but changes 1,145 correct predictions to incorrect ones. 
For Python, context increases GPT-5.4's FQN accuracy by 1.8 percentage points, with 294 incorrect predictions becoming correct and 214 correct predictions becoming incorrect. 
After Holm correction, the decreases in Java full signature accuracy for all four models and the increase in Python FQN accuracy for GPT-5.4 are statistically significant at $\alpha=0.05$, whereas the Python differences for the other three models are not. 
Overall, focused client call contexts can change recommendation outcomes, but the evaluated models do not benefit from it consistently. 

\textbf{Recommendation Performance by Mapping Frequency}. 
Figure~\ref{fig:rq4-mapping-frequency} shows a strong association between mapping frequency and recommendation accuracy. 
Under the signature input, the accuracy ranges from 15.4\% to 21.1\% for Java and from 23.9\% to 36.9\% for Python among mappings appearing in only one client update instance. 
For mappings appearing in more than 100 instances, accuracy increases to 48.5\%--55.8\% for Java and 73.8\%--79.4\% for Python. 
The context input exhibits a similar pattern. 
Across the evaluated models, languages, and input settings, recommendation accuracy is substantially higher for frequently observed mappings than for mappings observed in only a few client update instances. 
This result connects the long-tail distribution identified in RQ3 with a practical challenge for replacement API recommendation: many mappings lie in the long tail where recommendation accuracy is substantially lower. 

\begin{resultbox}
\textbf{Answer to RQ4:} 
The best FQN accuracy reaches only 37.1\% for Java and 44.4\% for Python, and focused client call context provides no consistent improvement. 
More importantly, all evaluated models achieve substantially lower accuracy on sparsely observed mappings than on recurrent mappings, highlighting their limitations in reliably identifying replacements for the long tail of API updates captured by our dataset. 
\end{resultbox}

\section{Discussion}\label{s: discussion}
\subsection{Implications}
\textbf{The \toolname dataset provides reusable information for developing and evaluating automated library update techniques.} 
RQ3 shows that the mined dataset connects API update mappings, version transitions, and client update instances across thousands of libraries. 
These three units provide complementary information about what API is replaced, between which library versions the replacement is observed, and how clients adapt the code. 
This information can support automated library update techniques in multiple ways. 
First, LLM-based techniques can retrieve mappings and associated client update instances from the dataset as replacement knowledge and real-world examples instead of relying solely on information manually supplied in prompts~\cite{GoogleCodeMigration, GUPPY, SQLAlchemy}. 
Second, the mappings and client update instances can provide examples for constructing transformation rules for recurring API changes~\cite{openrewrite, openrewriteMigrateLog4j}. 
Finally, since the dataset retains the originating dependency update commits, it can facilitate the construction of project-level benchmarks by identifying candidate projects and version transitions for further curation and build-based validation. 

\textbf{High-precision library update dataset construction benefits from combining client-side candidate generation with library-side validation.} 
RQ2 shows that similarity weights have limited impact on the F1 score, whereas library-side validation in Phase 4 increases precision by 40.3 percentage points for Java and 43.3 points for Python at the selected thresholds. 
This result supports the central design of \toolname: client code provides plausible replacement candidates, while library artifacts provide evidence for validating them. 
Mining approaches for related software evolution tasks may similarly benefit from using lightweight client-side signals for candidate generation and authoritative artifacts for subsequent validation rather than relying on increasingly complex similarity models alone. 
The matching threshold can then be selected according to the intended use: reusable dataset construction may favor precision, whereas interactive tools with developer review may tolerate lower precision in exchange for higher recall. 

\textbf{Replacement API recommendation remains challenging for current LLMs, particularly for mappings in the long tail.} 
RQ4 shows that the evaluated LLMs achieve limited accuracy in replacement API recommendation and do not consistently benefit from focused client call context. 
Because selecting an appropriate replacement API is a necessary step in library updates, these results suggest that replacement selection should be treated explicitly rather than assumed to be reliably handled by an LLM. 
A practical workflow therefore can separate replacement API selection from code adaptation: an update system can first identify which legacy APIs need to be replaced and select their replacement APIs, verify that the selected APIs are available in the target library version, and then adapt the affected client code. 
Compilation, type checking, and testing can subsequently validate the generated code changes. 
RQ4 further shows that recommendation accuracy is substantially lower for mappings observed in only a few client update instances than for frequently observed mappings. 
For mappings in this long tail, future LLM-based techniques may need to complement historical client examples with library-side information such as release notes, API documentation, and source code. 

\subsection{Threats to Validity}
\noindent\textbf{Internal Validity}. 
First, the ground truth dataset is sampled from API call change records produced in Phase 3 and therefore evaluates candidate matching and library-side validation rather than the end-to-end recall of the complete \toolname pipeline. 
Dependencies or API calls missed in earlier phases, as well as changes excluded during API call change detection, cannot enter the ground truth dataset. 
Therefore, the RQ1 recall should not be interpreted as the fraction of all client update instances in the studied ecosystems recovered by \toolname. 

Second, the collection and extraction process makes several assumptions to enable ecosystem-scale mining. 
\toolname assumes that dependency updates and their associated code adaptations occur within the same commit, although some co-occurring code changes may be unrelated to the dependency update. 
We mitigate this threat by restricting candidate API call changes to the updated library and validating the resulting candidates against the involved library versions. 
Phase 3 also uses the latest library artifact to identify import prefixes, assuming that these prefixes remain sufficiently stable across releases. 
Changes in import prefixes may cause API calls from earlier versions to be missed, reducing the coverage of candidate update instance mining. 
Moreover, the lightweight and conservative static analysis used to process millions of dependency configuration and code blobs may miss dependencies or API calls. 
Accordingly, the dataset characterized in RQ3 represents the library updates captured by \toolname rather than an exhaustive census of API updates in the Java and Python ecosystems. 

Third, candidate matching and library-side validation can introduce false positives and false negatives. 
Similarity-based matching may miss replacements with low similarity or pair similar but unrelated API calls. 
Although Phase 4 substantially improves precision, removal or deprecation of a legacy API does not establish that the paired replacement API is the intended replacement. 
Conversely, API signature resolution failures and unrecognized deprecation mechanisms can discard genuine update instances. 
The RQ1 error analysis characterizes these cases, and the evaluation shows precision above 90\% for both languages. 
Nevertheless, applications requiring manually verified replacement relations should account for the remaining errors in the automatically mined dataset. 

Finally, manual annotation in the ground truth onstruction and the reference replacements used in RQ4 may contain uncertainty. 
To reduce annotation errors, two authors independently annotated all possible API call pairs using external evidence, resolved disagreements through discussion, and achieved a Cohen's $\kappa$ of 0.92. 
For RQ4, the reference mappings are automatically mined by \toolname and therefore have high but imperfect precision. 
To increase the reliability of the reference replacements, we retain a replacement only when it accounts for a strict majority of the client update instances for the same library, version transition, and legacy API. 
Besides, the same legacy API may have multiple valid replacements for a version transition. 
Since each RQ4 example uses only one reference replacement, exact-match accuracy may treat other valid replacements as incorrect. 
RQ4 should therefore be interpreted as evaluating recommendation against the replacement relations captured by the dataset rather than all potentially valid replacements. 

\noindent\textbf{External Validity}. 
First, the current implementation of \toolname covers only a subset of dependency updates. 
It targets pinned dependency updates declared through Maven POM files and four common Python dependency files and requires the dependency version change and associated code changes to occur in the same commit. 
Consequently, the resulting dataset does not cover updates expressed through unsupported dependency configuration files, unpinned version constraints, or client adaptations distributed across multiple commits. 
The dataset is constructed from World of Code V3 and therefore reflects the projects and development history captured by this snapshot. 
Its scale and distribution characteristics may differ for newer snapshots or other software repositories. 
Moreover, \toolname currently supports only Java and Python. 
Applying the framework to other ecosystems requires implementing the language-dependent components in Figure~\ref{fig: overview} according to their packaging practices, language features, and deprecation mechanisms. 

Second, the ground truth dataset covers 100 sampled libraries per language, and the mining effectiveness of \toolname may differ for libraries and APIs outside this sample. 
Our two-stage sampling strategy promotes diversity across libraries and old-side APIs rather than allowing frequently updated libraries or APIs to dominate the evaluation. 
Therefore, the reported precision, recall, and F1 scores should not be interpreted as population-weighted estimates over all API call change records. 

Finally, RQ4 evaluates four LLMs using one prompt template and two input settings. 
The recommendation accuracy may therefore differ for other models, prompts, or input settings. 
The association between mapping frequency and recommendation accuracy should be interpreted as an empirical finding for the evaluated models and dataset rather than as a causal relationship. 
Mapping frequency may correlate with other factors, such as API popularity and the availability of documentation and code examples. 

\section{Conclusion}\label{s: conclusion}
We present \toolname, a client-driven framework for automatically constructing ecosystem-scale library update datasets from client dependency updates. 
\toolname mines candidate update instances from client-side code changes, validates the candidates against involved library versions, and derives API update mappings from the validated instances. 
This design connects API update mappings, version transitions, and client update instances while grounding each retained mapping in at least one client update instance. 
On a manually annotated ground truth dataset, \toolname achieves precision of 91.6\% for Java and 90.1\% for Python, with library-side validation improving precision by more than 40 percentage points for both languages. 
Applied to World of Code V3, \toolname mines 381,661 Java and 277,259 Python client update instances, representing 18,900 and 4,456 API update mappings across 2,557 and 999 libraries, respectively. 
The mined mappings exhibit a pronounced long-tail distribution, with most appearing in only a few client update instances and a small number appearing substantially more frequently. 
Using replacement API recommendation as one application of the dataset, we find that the four evaluated LLMs achieve limited accuracy and perform substantially better on frequently observed mappings than on mappings observed in only a few client update instances. 
This result highlights replacement API recommendation for mappings in the long tail as a challenge for current LLMs. 

Overall, \toolname provides an automated approach to constructing ecosystem-scale library update datasets that connect what APIs are replaced, between which library versions, and how clients perform the updates. 
In future work, we plan to extend \toolname to additional dependency configurations and programming language ecosystems and investigate how the mined mappings and client update instances can support automated library updates. 

\begin{acks}
This work is supported by the National Natural Science Foundation of China (62502030, 92582119, 62272037), China Postdoctoral Science Foundation (2025M781455), and Fundamental Research Funds for the Central Universities (No. FRF-TP-25-030).
\end{acks}

\bibliographystyle{ACM-Reference-Format}
\bibliography{references}

\end{document}